\documentstyle[12pt]{article}
\renewcommand{\thefootnote}{\fnsymbol{footnote}}

\begin{document}
\baselineskip 6mm
\begin{flushright}
  {\tt TMI-97-1 \\
      March 1997}
\end{flushright}
\thispagestyle{empty}
\vspace*{5mm}
\begin{center}
   {\Large {\bf Fermion-Boson-Type Subquark Model and 
           ${\Delta}F=2$ Phenomena}}
\vspace*{20mm} \\
  {\sc Takeo~~Matsushima}
 \footnote[1]
  {
   $\dagger$ Available mail address is as follows :
\par
     Nagoya University Department of Physics 
     Chikusa-ku Nagoya 464-01 Japan
\par
   e-mail : mtakeo@eken.phys.nagoya-u.ac.jp }
\vspace{5mm} \\
  \sl{17-1 1-Chome Azura Ichinomiya,}\\
  \it{Aichi-Prefecture 491, Japan}

\vspace{15mm}  

\center{\bf Abstract} \\
\vspace{3mm}
\begin{minipage}[t]{120mm}
\baselineskip 5mm
{\small
 ${\Delta}F=2$ phenomena in all neutral pseudo scalar meson 
 systems are investigated in the context of the 
 fermion-boson-type 
 subquark model. In this model the mass difference between 
 the heavier neutral pseudo scalar meson and the lighter 
 one and indirect CP violation associated with the 
 mixing are unifyingly explained by the neutral scalar 
 subquark exchanges between two quarks inside the present 
 meson. We obtain 
 the mass differences : ${\Delta}M_{D}\approx10^{-14}$ GeV, 
 ${\Delta}M_{B_s}\approx10^{-11}$ GeV, ${\Delta}M_{T_u}
 \approx10^{(-10\sim{-9})}$ GeV and ${\Delta}M_{T_c}
 \approx10^{(-8\sim{-7})}$ GeV. It is also predicted that 
 indirect CP violations in D- and B-meson systems are same
 as in K-meson system.
 }
\end{minipage}
\end{center}



\newpage
\baselineskip 18pt
\section{Introduction}
\hspace*{\parindent}
 Since the first observation of CP violation in the $K^0$ 
 system in 1964[1], this evidence has not been explained 
 in a successful scenario. 
 It was soon after this experiment in the same year 
 that Wolfenstein proposed the idea of super-weak theory 
 that CP violation occurs only in the effective 
 Hamiltonian which describes the time evolution of the 
 $K^0$ system and its strength is $10^{-9}$ times smaller 
 than the standard weak interactions. 
 In 1973 Kobayashi and Maskawa introduced the $3\times3$ 
 unitary matrix which mixes three generations in the 
 context of the electroweak unified gauge theory called 
 the standard model (SM) usually. 
 This model has the ability to explain CP violations 
 not only in the effective Hamiltonian of the second 
 order weak interaction but also in the decay amplitudes 
 if two different decay processes contribute to one final 
 state. 
 In the latter case the parameter 
 $Re({\epsilon}^{'}/{\epsilon})_{K}\neq0$ is expected 
 although a large top quark mass effect gives 
 the prediction of very small value[3]. On the other 
 hand the super-weak theory predict the exact zero value.  
 The experimental efforts about 
 $Re({\epsilon}^{'}/{\epsilon})_K$
 so far have not given a conclusive answer. 
 Recently CP violations in heavier neutral mesons 
 ($D^0$, $B^0$) have been discussed 
 as the concerning experiments were carried out (though 
 not yet been observed) and in the near future the real 
 features will become clear. 
 So the totally understandable scenario of CP violation 
 phenomena is required. The mass difference (${\Delta}M_P$) 
 between heavier neutral pseudo scalar ($P$-) meson and 
 lighter one is considered to be essentially connected 
 with CP violations 
 because the origin of both phenomena comes 
 from the off diagonal matrix elements of the mass 
 matrix $(M_{ij},i,j=1,2)$ and the decay matrix
 (${\Gamma}_{ij},i,j=1,2$).
 Comparing with CP violations, the experiments of 
 ${\Delta}M_P$ are a little abundant, e.g., there are 
 ${\Delta}M_K$, ${\Delta}M_{B_d}$, the upper bound of 
 ${\Delta}M_D$, and the lower bound of ${\Delta}M_{B_s}$[4]. 
 Theoretical analyses about them are roughly in two ways, 
 e.g., the estimation of 
 $M_{12}$ by the superweak theory or the box diagram 
 calculation in the SM with (or without) long distance 
 contributions. Therefore the aim of the present stage 
 is to clarify what kind of dynamics controls $M_{ij}$ and 
 ${\Gamma}_{ij}$. In this paper we investigate this issue 
 in the context of the Fermion-Boson-type subquark 
 model (FB-model) inspired by some type of gauge 
 theory which proposed by the author[5]. In this model 
 the assumption that the ``neutral scalar subquarks 
 (${\bf y}$) exchange'' between two quarks inside the 
 present $P$-meson plays the essential role. 
 So this scenario may be said a realization of 
 Wolfenstein's super-weak idea[2]. 
 In Sect.2 we mention the gauge theory 
 which inspires quark-lepton composite scenario. 
 In Sect.3 we review the FB-model by which we study the 
 present issue. In Sect.4 we investigate the mass 
 differences and CP violations in neutral pseudo scalar 
 meson systems by using this FB-model.

\section{Gauge theory inspiring quark-lepton composite 
 scenario}
\hspace*{\parindent}
  In our model the existence of fundamental matter fields 
 (preon) are inspired by the gauge theory with Cartan 
 connections[5]. 
 Let us briefly summarize the basic features of that. 
 Generally gauge fields, including gravity, are considered 
 as geometrical objects, that is, connection coefficients 
 of principal fiber bundles. 
 It is said that there exist some different points 
 between Yang-Mills gauge theories and gravity, 
 though both theories commonly possess fiber bundle 
 structures. The latter has the fiber bundle related 
 essentially to 4-dimensional space-time freedoms 
 but the former is given, in an ad hoc way, the one with 
 the internal space which has nothing to do with 
 the space-time coordinates. 
 In case of gravity it is usually considered that 
 there exist ten gauge fields, that is, six spin 
 connection fields in $SO(1,3)$ gauge group and four 
 vierbein fields in $GL(4,R)$ gauge group from which the 
 metric tensor ${\bf g}^{{\mu}{\nu}}$ is constructed 
 in a bilinear function of them. Both altogether belong to 
 Poincar\'e group $ISO(1,3)=SO(1,3)\otimes{R}^4$ 
 which is semi-direct product. 
 In this scheme spin connection 
 fields and vierbein fields are independent but only if 
 there is no torsion, both come to have some relationship. 
 Seeing this, $ISO(1,3)$ gauge group theory has the logical 
 weak point not to answer how two kinds of gravity fields 
 are related to each other intrinsically.
\par   
In the theory of Differential Geometry, S.Kobayashi has 
 investigated the theory of ``Cartan connection''[6]. 
 This theory, in fact, has ability to reinforce the above 
 weak point. 
 The brief recapitulation is as follows. 
 Let $E(B_n,F,G,P)$ be a fiber bundle (which we call 
 Cartan-type bundle) associated with a principal fiber 
 bundle $P(B_n,G)$ where $B_n$ is a base manifold with  
 dimension ``$n$'', $G$ is a structure group, $F$ is a 
 fiber space which is homogeneous and diffeomorphic with 
 $G/G'$ where $G'$ is a subgroup of $G$. 
 Let $P'=P'(B_n,G')$ be a principal fiber bundle, 
 then $P'$ is a subbundle of $P$.  
 Here let it be possible to decompose the Lie algebra 
 ${\bf g}$ of $G$ into the subalgebra ${\bf g}'$ of $G'$ 
 and a vector space ${\bf f}$ such as : 
\begin{equation}
 {\bf g}=
 {\bf g}'+{\bf f},\hspace{1cm}{\bf g}'\cap{\bf f}=0,
 \label{1}
\end{equation}
\begin{equation}
 [{\bf g'},{\bf g'}]\subset {\bf g'},\label{2}
\end{equation}
\begin{equation}
 [{\bf g'},{\bf f}]\subset {\bf f},\label{3}
\end{equation}
\begin{equation}
 [{\bf f},{\bf f}]\subset {\bf g'},\label{4}  
\end{equation}
 where $dim{\bf f}=dimF=dimG-dimG'=dimB_n=n$. 
 The homogeneous space $F=G/G'$ is said to be 
 ``weakly reductive'' if there exists a vector space 
 ${\bf f}$ satisfying Eq.(1) and (3). 
 Further $F$ satisfying Eq(4) is called ``symmetric space''. 
 Let ${\bf \omega}$ denote the connection form of $P$ and 
 $\overline{\bf \omega}$ be the restriction of 
 ${\bf \omega}$ to $P'$. Then $\overline{\bf \omega}$ is 
 a ${\bf g}$-valued linear differential 1-form 
 and we have :
\begin{equation}
 {\bf \omega}=
 g^{-1}\overline{\bf \omega}g+g^{-1}dg,\label{5}
\end{equation}
 where $g\in{G}$, $dg\in{T_g(G)}$. ${\bf \omega}$ is 
 called the form of ``Cartan connection'' in $P$. 
\par
 Let the homogeneous space $F=G/G'$ be weakly reductive. 
 The tangent space $T_O(F)$ at $o\in{F}$ is isomorphic 
 with ${\bf f}$ and then $T_O(F)$ can be identified with 
 ${\bf f}$ and also 
 there exists a linear ${\bf f}$-valued differential 
 1-form(denoted by ${\bf \theta}$) which we call the 
 ``form of soldering''. Let ${\bf \omega}'$ 
 denote a ${\bf g}'$-valued 1-form in $P'$, we have :
\begin{equation}
\overline{\bf \omega}={\bf \omega}'+{\bf \theta}.\label{6}
\end{equation}
 The dimension of vector space ${\bf f}$ and the dimension 
 of base manifold $B_n$ is the same ``$n$'', 
 and then ${\bf f}$ can be identified with the tangent 
 space of $B_n$ at the same point in $B_n$ and 
 ${\bf \theta}$s work as $n$-bein fields. 
 In this case  ${\bf \omega}'$ and ${\bf \theta}$ 
 unifyingly belong to group $G$. 
 Here let us call such a mechanism ``Soldering Mechanism''.
\par
 Drechsler has found out the useful aspects of this theory
  and 
 investigated a gravitational gauge theory based on 
 the concept of the Cartan-type bundle equipped with the 
 Soldering Mechanism[7]. He considered $F=SO(1,4)/SO(1,3)$ 
 model. Homogeneous space $F$ with $dim=4$ solders 
 4-dimensional real space-time. 
 The Lie algebra of $SO(1,4)$ corresponds to 
 ${\bf g}$ in Eq.(1), that of $SO(1,3)$ corresponds to 
 ${\bf g}'$ and ${\bf f}$ is 4-dimensional vector space. 
 The 6-dimensional spin connection fields are 
 ${\bf g}'$-valued objects and vierbein fields are 
 ${\bf f}$-valued, both of which are unified into 
 the members of $SO(1,4)$ gauge group. 
 We can make the metric tensor ${\bf g}^{{\mu}{\nu}}$ as 
 a bilinear function of ${\bf f}$-valued vierbein fields. 
 Inheriting Drechsler's study the author has investigated 
 the quantum theory of gravity[5]. 
 The key point for this purpose is that $F$ is a symmetric 
 space because ${\bf f}$s are satisfied with Eq.(4). 
 Using this symmetric nature we can inquire into making a 
 quantum gauge theory, that is, constructing 
 ${\bf g}'$-valued Faddeev-Popov ghost, anti-ghost, 
 gauge fixing, gaugeon and its pair field as composite 
 fusion fields of ${\bf f}$-valued gauge fields by use 
 of Eq.(4) and also naturally inducing BRS-invariance. 
\par
 Comparing such a scheme of gravity, let us consider 
 Yang-Mills gauge theories. Usually when we make the 
 Lagrangian density 
 ${\cal L}=tr({\bf F}\wedge{\bf F}^{\ast})$ 
 (${\bf F}$ is a field strength), we must borrow 
 a metric tensor ${\bf g}^{{\mu}{\nu}}$ from gravity to 
 get ${\bf F}^{\ast}$ and also for Yang-Mills gauge 
 fields to propagate in the 4-dimensional real space-time. 
 This seems to mean 
 that ``there is a hierarchy between gravity and other three 
 gauge fields (electromagnetic, strong, and weak)''. But is 
 it really the case ? 
 As an alternative thought we can think that all kinds of 
 gauge fields are ``equal''. 
 Then it would be natural for the question 
 ``What kind of equality is that ?'' to arise. 
 In other words, it is the question that 
 ``What is the minimum structure of the gauge mechanism 
 which four kinds of forces are commonly equipped with ?''. 
 For answering this question, 
 let us make a assumption : ``\begin{em}Gauge fields 
 are Cartan connections equipped with Soldering 
 Mechanism\end{em}.'' 
 In this meaning all gauge fields are equal. 
 If it is the case three gauge fields except for gravity 
 are also able to have their own metric tensors and to 
 propagate in the real space-time without the help of 
 gravity. Such a model has already investigated in Ref.[5]. 
\par
 Let us discuss them briefly. It is found that there 
 are four types of sets of classical groups with smaller 
 dimensions which admit Eq.(1,2,3,4), that is, 
 $F=SO(1,4)/SO(1,3)$, $SU(3)/U(2)$, $SL(2,C)/GL(1,C)$ 
 and $SO(5)/SO(4)$ with $dimF=4$[8]. 
 Note that the quality of ``$dim\hspace{1mm}4$''
 is very important because it guarantees $F$ to solder 
 to 4-dimensional real space-time and all gauge fields 
 to work in it. 
 The model of $F=SO(1,4)/SO(1,3)$ for gravity 
 is already mentioned. 
 Concerning other gauge fields, it seems to be 
 appropriate to assign $F=SU(3)/U(2)$ to QCD gauge 
 fields, $F=SL(2,C)/GL(1,C)$ to QED gauge fields and 
 $F=SO(5)/SO(4)$ to weak interacting gauge fields. 
 Some discussions concerned are following. 
 In general, matter fields couple to ${\bf g}'$-valued 
 gauge fields. As for QCD, matter fields couple to 
 the gauge fields of $U(2)$ subgroup but $SU(3)$ contains, 
 as is well known, three types of $SU(2)$ subgroups and 
 then after all they couple to all members of $SU(3)$
  gauge fields. 
 In case of QED, $GL(1,C)$ is locally isomorphic with 
 $C^1\cong{U(1)}\otimes{R}$. 
 Then usual Abelian gauge fields are assigned to $U(1)$ 
 subgroup of $GL(1,C)$. Georgi and Glashow suggested that  
 the reason why the electric charge is quantized comes 
 from the fact that $U(1)$ electromagnetic gauge group 
 is a unfactorized subgroup of $SU(5)$[9]. 
 Our model is in the same situation because $GL(1,C)$ 
 a unfactorized subgroup of $SL(2,C)$. 
 For usual electromagnetic $U(1)$ gauge group, the electric 
 charge unit ``$e$''$(e>0)$ is for $one\hspace{2mm} 
 generator$ of $U(1)$ but in case of $SL(2,C)$ 
 which has $six\hspace{2mm} generators$, the minimal 
 unit of electric charge shared per one generator must be 
 ``$e/6$''. 
 This suggests that quarks and leptons might have the 
 substructure simply because $e,\hspace{1mm}2e/3,
 \hspace{1mm}e/3>e/6$.
  Finally as for weak interactions we adopt 
 $F=SO(5)/SO(4)$. It is well known that $SO(4)$ is 
 locally isomorphic with $SU(2)\otimes{SU(2)}$. 
 Therefore 
 it is reasonable to think it the left-right symmetric 
 gauge group : $SU(2)_L\otimes{SU(2)}_R$. 
 As two $SU(2)$s are direct product, it is able to have 
 coupling constants (${\bf g}_L,{\bf g}_R$) independently. 
 This is convenient to explain the fact of the 
 disappearance of right-handed weak interactions 
 in the low-energy region. 
 Possibility of composite structure of quarks and leptons 
 suggested by above $SL(2,C)$-QED would introduce 
 the thought that the usual left-handed weak interactions 
 are intermediated by massive composite vector bosons as 
 ${\bf \rho}$-meson in QCD and that they are residual 
 interactions due to substructure dynamics of quarks 
 and leptons. The elementary massless gauge fields 
 relate essentially to the structure of the real space-time  
 manifold as the connection fields but on the other hand 
 the composite vector bosons have nothing to do with it. 
 Considering these discussions, we shall set the 
 assumption : ``\begin{em}All kinds of gauge fields are 
 elementary massless fields, belonging to spontaneously 
 unbroken $SU(3)_C\otimes{SU(2)}_L\otimes{SU(2)}_R
 \otimes{U(1)}_{e.m}$ gauge group and quarks 
 and leptons and {\bf W, Z} are all composite objects
 of the elementary matter fields\end{em}.''

\section{Composite model}
\hspace*{\parindent}
 Our direct motivation towards compositeness of quarks 
 and leptons is one of the results of the arguments 
 in Sect.2, that is, 
 $e,\hspace{1mm}2e/3,\hspace{1mm}e/3>e/6$. 
 However, other several phenomenological 
 facts tempt us to consider a composite model, e.g.,  
 repetition of generations, quark-lepton parallelism of 
 weak isospin doublet structure, 
 quark-flavor-mixings, etc. .
 Especially Bjorken[10]'s and Hung and Sakurai[11]'s 
 suggestion of an alternative to unified 
 weak-electromagnetic gauge theories have invoked many 
 studies of composite models including 
 composite weak bosons[12-17]. 
 Our model is in the line of those studies. 
 There are two ways to make composite models, that is, 
 ``Preons are all fermions.'' or ``Preons are   
  both fermions and bosons (FB-model).'' 
 The merit of the former is that it can avoid the problem 
 of a quadratically divergent self-mass of elementary 
 scalar fields. 
 However, even in the latter case such a disease 
 is overcome if both fermions and bosons are the 
 supersymmetric pairs, both of which carry the same 
 quantum numbers except for the nature of Lorentz 
 transformation (spin-1/2 or spin-0)[18]. 
 Pati and Salam have suggested    
 that the construction of a neutral composite object 
 (neutrino in practice) needs both kinds of preons, 
 fermionic as well as bosonic, 
 if they carry the same charge for the Abelian gauge or 
 belong to the same (fundamental) representation for the 
 non-Abelian gauge[19].  
 This is a very attractive idea for constructing 
 the minimal model. 
 Further, from the representation theory of Poincar\'e 
 group both integer and half-integer spin angular 
 momentum occur equally for massless particles[20]. 
 If nature chooses ``fermionic monism'', 
 there must exist the additional special reason 
 to select it. 
 Then in this point also, the thought of the FB-model 
 is minimal. 
 Based on such considerations we proposed a FB-model of 
 \begin{em}``only one kind of spin-1/2 elementary field 
 ($\Lambda$) and of spin-0 elementary field 
 ($\Theta$)''\end{em}[5]. 
 Both have the same electric charge of ``$e/6$''  
 (Maki has first proposed the FB-model with the universal 
 electric charge $e/6$. \cite{21})
\renewcommand{\thefootnote}{\arabic{footnote}}
\footnote{The notations of $\Lambda$ and $\Theta$ are 
 inherited from those in Ref.[21]. 
 After this we shall call $\Lambda$ and $\Theta$ 
 ``{\bf Primon}'' named by Maki which means 
 ``primordial particle''.}
 and the same transformation properties of the 
 fundamental representation ( 3, 2, 2) under the 
 spontaneously unbroken gauge symmetry of 
 $SU(3)_C\otimes{SU(2)_L}\otimes{SU(2)_R}$
 (let us call $SU(2)_L\otimes{SU(2)_R}$ ``hypercolor 
 gauge symmetry''). 
 Then $\Lambda$ and $\Theta$ come into the supersymmetric 
 pair which guarantees 'tHooft's naturalness condition[22]. 
 The $SU(3)_C$, $SU(2)_L$ and $SU(2)_R$ gauge fields 
 cause the confining forces with confining energy scales of 
 $\Lambda_c<< \Lambda_L<(or \cong) \Lambda_R$ (Schrempp and 
 Schrempp discussed them elaborately in Ref.[17]). 
 Here we call positive-charged primons ($\Lambda$, $\Theta$) 
 ``$matter$'' and negative-charged primons 
 ($\overline\Lambda$, $\overline\Theta$) ``$antimatter$''. 
 Our final goal is to build quarks, leptons and ${\bf W, Z}$ 
 from primons : $\Lambda$ ($\overline\Lambda$)
  and $\Theta$ ($\overline\Theta$).   
 Let us discuss that scenario next.                                             \par
 At the very early stage of the development of the universe, 
 the matter fields ($\Lambda$, $\Theta$) and their 
 antimatter fields ($\overline{\Lambda}$, 
 $\overline{\Theta}$) must have broken out from the vaccum. 
 After that they would have combined with each  
 other as the universe was expanding. That would be the 
 first step of the existence of composite matters. 
 There are ten types of them : 
\newline
 \hspace*{2mm}$spin1/2$\hspace{2cm}$spin0$\hspace{2.7cm}
 $e.m.charge$\hspace{1.2cm}$Y.M.representation$
{
\setcounter{enumi}{\value{equation}}
\addtocounter{enumi}{1}
\setcounter{equation}{0} 
\renewcommand{\theequation}{\theenumi\alph{equation}}
\begin{eqnarray}
 \Lambda\Theta\hspace{2.5cm}\Lambda\Lambda,
 \Theta\Theta\hspace{3.1cm}
 e/3\hspace{1.8cm}(\overline{3},1,1)\hspace{2mm}
 (\overline{3},3,1)\hspace{2mm}(\overline{3},1,3),\\
 \Lambda\overline\Theta,\overline\Lambda\Theta
 \hspace{2cm}\Lambda\overline\Lambda,\Theta\overline
 \Theta\hspace{3.2cm}0\hspace{2.1cm}(1,1,1)\hspace{2mm}
 (1,3,1)\hspace{2mm}(1,1,3),\\
 \overline\Lambda\overline\Theta\hspace{2.5cm}
 \overline\Lambda\overline\Lambda,\overline\Theta
 \overline\Theta\hspace{2.7cm}-e/3\hspace{1.7cm}(3,1,1)
 \hspace{2mm}(3,3,1)\hspace{2mm}(3,1,3)\hspace{1mm}.
 \label{7}
\end{eqnarray}
\setcounter{equation}{\value{enumi}}}
 In this step the confining forces are, in kind, 
 in $SU(3)\otimes{SU(2)}_L\otimes{SU(2)}_R$ gauge symmetry 
 but the $SU(2)_L\otimes{SU(2)}_R$ confining forces must 
 be main because of the energy scale of 
 $\Lambda_L,\Lambda_R>>\Lambda_c$
 and then the color gauge coupling $\alpha_s$ 
 and e.m. coupling constant $\alpha$ are negligible. 
 As is well known, the coupling constant of $SU(2)$ 
 confining force are characterized by 
 $\varepsilon_i=\sum_a\sigma_\alpha^a\sigma_\beta^a$,where 
 ${\sigma}s$ are $2\times2$ matrices of $SU(2)$, $a=1,2,3$, 
 $\alpha,\beta=\Lambda,\overline\Lambda,\Theta,\overline
 \Theta$, $i=0$ for singlet and $i=3$ for triplet. 
 They are calculated as $\varepsilon_0=-3/4$ which causes 
 the attractive force and and $\varepsilon_3=1/4$ causing 
 the repulsive force. 
 As concerns, $SU(3)_C$ octet and sextet states are 
 repulsive but singlet, triplet and antitriplet states are
  attractive and then the formers are disregarded. 
 Like this, two primons are confined into composite objects 
 with more than  one singlet state of any $SU(3)_C\otimes 
 {SU(2)}_L\otimes{SU(2)}_R$. Note that three primon systems 
 cannot make the singlet states of $SU(2)$. 
 Then we omit them. In Eq.(7,b), the $(1,1,1)$-state is the 
 ``most attractive channel''. 
 Therefore $(\Lambda\overline\Theta)$, $(\overline\Lambda
 \Theta)$, $(\Lambda\overline\Lambda)$ and 
 $(\Theta\overline\Theta)$ of $(1,1,1)$-states 
 with neutral e.m. charge must 
 have been most abundant in the universe. 
 Further $(\overline{3},1,1)$- and $(3,1,1)$-states 
 in Eq.(7,a,c) are next attractive. 
 They presumably go into $\{(\Lambda\Theta)
 (\overline\Lambda\overline\Theta)\}, 
 \{(\Lambda\Lambda)(\overline\Lambda\overline\Lambda)\}$, 
 etc. of $(1,1,1)$-states with neutral e.m. charge.
 These objects may be the candidates for the 
 ``cold dark matters'' if they have even tiny masses. 
 It is presumable that the ratio of the quantities between 
 the ordinary matters and the dark matters firstly depends 
 on the color and hypercolor charges (maybe the ratio is 
 more than $1/3\times3$). Finally the $(*,3,1)$- and 
 $(*,1,3)$-states are remained ($*$ is $1,3,\overline{3}$). 
 They are also stable
  because $|\varepsilon_0|>|\varepsilon_3|$. 
 They are, so to say, the ``intermediate clusters'' towards 
 constructing ordinary matters, 
\footnote{Such thoughts have been proposed by Maki 
 in Ref.[21]} namely quarks, leptons and ${\bf W,Z}$. 
 Here we call such intermediate clusters ``subquarks'' and 
 denote them as follows :
\newline
\hspace*{6.5cm}$Y.M.representation$\hspace{1.5cm}$spin$
 \hspace{0.5cm}$e.m.charge$
{
\setcounter{enumi}{\value{equation}}
\addtocounter{enumi}{1}
\setcounter{equation}{0} 
\renewcommand{\theequation}{\theenumi\alph{equation}}
\begin{eqnarray}
 {\bf \alpha}&=&
 (\Lambda\Theta),\hspace{2.2cm}{\bf \alpha}_L:
 (\overline{3},3,1),\hspace{3mm}{\bf \alpha}_R:
 (\overline{3},1,3)\hspace{1.2cm}1/2\hspace{1.2cm}e/3,\\
 {\bf \beta}&=&
 (\Lambda\overline\Theta),\hspace{2.2cm}{\bf \beta}_L: 
 (1,3,1),\hspace{3mm}{\bf \beta}_R:(1,1,3)
 \hspace{1.2cm}1/2\hspace{1.5cm}0,\\
 {\bf x}&=&
 (\Lambda\Lambda,\hspace{2mm}\Theta\Theta),\hspace{1.2cm}
 {\bf x}_L:(\overline{3},3,1),\hspace{3mm}{\bf x}_R:
 (\overline{3},1,3)\hspace{1.3cm}0\hspace{1.5cm}e/3,\\
 {\bf y}&=&
 (\Lambda\overline\Lambda,\hspace{2mm}\Theta
 \overline\Theta),\hspace{1.2cm}{\bf y}_L:(1,3,1),
 \hspace{3mm}{\bf y}_R:(1,1,3)\hspace{1.3cm}0
 \hspace{1.7cm}0,\label{8}
\end{eqnarray}
\setcounter{equation}{\value{enumi}}}
 and there are also their antisubquarks[16].
\footnote{The notations of ${\bf \alpha}$,${\bf \beta}$,
 ${\bf x}$ and ${\bf y}$ are inherited from those 
 in Ref.[16] written by Fritzsch and Mandelbaum, 
 because ours is, in the subquark level, similar to 
 theirs with two fermions and two bosons.} 
\par
 Now we come to the step to build quarks and leptons. 
 The gauge symmetry of the confining forces in this step 
 is also $SU(2)_L\otimes{SU(2)}_R$ 
 because the subquarks are in the triplet states of 
 $SU(2)_{L,R}$ and then they are combined into singlet 
 states by the decomposition of $3\times3=1+3+5$
 in $SU(2)$. We make the first generation as follows :
\newline
\hspace*{8cm}$e.m.charge$\hspace{1.2cm}$Y.M.representation$
{
\setcounter{enumi}{\value{equation}}
\addtocounter{enumi}{1}
\setcounter{equation}{0}
\renewcommand{\theequation}{\theenumi\alph{equation}} 
\begin{eqnarray}
 <{\bf u}_i|&=&
 <{\bf \alpha}_i{\bf x}_i|\hspace{3.4cm}2e/3\hspace{3cm}
 (3,1,1),\\
 <{\bf d}_i|&=&
 <\overline{\bf \alpha}_i\overline{\bf x}_i{\bf x}_i|
 \hspace{2.7cm}-e/3\hspace{3cm}(3,1,1),\\
 <{\bf \nu}_i|&=&
 <{\bf \alpha}_i\overline{\bf x}_i|\hspace{3.5cm}0
 \hspace{3.5cm}(1,1,1),\\
 <{\bf e}_i|&=&
 <\overline{\bf \alpha}_i\overline{\bf x}_i
 \overline{\bf x}_i|\hspace{2.7cm}-e\hspace{3.4cm}
 (1,1,1),\label{9}
\end{eqnarray}
\setcounter{equation}{\value{enumi}}}
 where $i$ stands for $L$ or $R$[12].
\footnote{Subquark configurations in Eq.(9) are 
 essentially the same as those in Ref.[12] written by 
 Kr\H olikowski, who proposed the model of one fermion and
 one boson with the same charge $e/3$.}
 Here we note that ${\bf \beta}$ and ${\bf y}$ 
 do not appear.  
 In practice ($({\bf \beta}{\bf y}):(1,1,1)$)-particle 
 is a candidate for neutrino. But as Bjorken has pointed
 out[10], non-vanishing charge radius of neutrino is 
 necessary for obtaining the correct low-energy effective 
 weak interaction Lagrangian[17]. 
 Therefore ${\bf \beta}$ is assumed not to contribute 
 to forming quarks and leptons. 
 Presumably composite
 (${\bf \beta}$${\bf \beta}$)-;(${\bf \beta}
 \overline{\bf \beta}$)-;($\overline{\bf \beta}
 \overline{\bf \beta}$)-states may go into the dark matters. 
 It is also noticeable that in this model the leptons have 
 finite color charge radius and then $SU(3)$ gluons interact 
 directly with the leptons at energies of the order of, 
 or larger than $\Lambda_{L}$ or $\Lambda_{R}$[18]. 
 Concerning the confinements of primons and subquarks, 
 the  confining forces of two steps are in the same 
 spontaneously unbroken $SU(2)_L\otimes{SU(2)}_R$ 
 gauge symmetry. Here let us assume that subquarks in 
 quarks are confined at the energy of $1.6$ TeV 
 (if admitting CDF's data[23]). 
\par
 Concerning the running coupling constant of the $SU(2)$ 
 gauge theory ($\alpha_{W}(Q^2)$) we know the following 
 equation :
{
\setcounter{enumi}{\value{equation}}
\addtocounter{enumi}{1}
\setcounter{equation}{0}
\renewcommand{\theequation}{\theenumi\alph{equation}}
\begin{eqnarray}
 1/\alpha_{W}(Q^2)&=&
 b_{2}ln(Q/{\Lambda}_q)^2,\hspace{3.5cm}\\
 b_{2}\hspace{0.7cm}&=&
 1/(4\pi)\{22/3-(2/3){N}_{f}-(1/12){N}_{s}\},\label{10}
\end{eqnarray}
\setcounter{equation}{\value{enumi}}}
 where $Q$ is the effective energy of $SU(2)$ gluon 
 exchange, ${\Lambda}_q$ is the confinement scale of 
 subquarks inside quarks and $N_{f}$($N_{s}$) is the 
 numbers of fermions (scalars) contributing to the 
 vacuum polarizations. 
 We calculate $b_2=0.35$ which  comes from that the 
 number of confined fermionic subquarks are $4$ 
 (${\bf \alpha}_{i},i=1,2,3$ for color freedom, 
 ${\bf \beta}$) and $4$ for bosons (${\bf x}_i, 
 {\bf y}$) contributing to the vacuum polarization. 
 Using $b_2=0.35$ we get $\alpha_{W}=0.040$ 
 at Q=$10^{19}$ GeV and extrapolating from this value 
 we obtain the result that the confining energy of primons 
 ($\Lambda$,$\Theta)$ is $1.6\times{10}^2$ TeV, 
 where we use $b_2=0.41$ which is calculated with three 
 kinds of $\Lambda$ and $\Theta$ owing to 
 three color freedoms.
 In sum, the radii of ${\bf \alpha}$, ${\bf \beta}$, 
 ${\bf x}$ and ${\bf y}$ are the inverse of $1.6
 \times{10}^2$ TeV and the radii of quarks are the 
 inverse of $1.6$ TeV. 
\par
 Next let us see the higher generations. 
 Harari and Seiberg have stated that the orbital and 
 radial excitations seem to have the wrong energy scale 
 ( order of $\Lambda_{L,R}$) and then the most likely 
 type of excitations is the addition of preon-antipreon 
 pairs[13,24].
 Then using ${\bf y}_{L,R}$ in Eq.(8,d) 
 we construct them as follows :
{  
\setcounter{enumi}{\value{equation}}
\addtocounter{enumi}{1} 
\setcounter{equation}{0}
\renewcommand{\theequation}{\theenumi\alph{equation}}
\begin{eqnarray}
&&\left\{
\begin{array}{lcl}
 <{\bf c}|&=&<{\bf \alpha}{\bf x}{\bf y}|\\
 <{\bf s}|&=&<\overline{\bf \alpha}\overline{\bf x}
{\bf x}{\bf y}|,
\end{array} 
\right.
\hspace{6mm}
\left\{
\begin{array}{lcl}
 <{\bf \nu_\mu}|&=&
 <{\bf \alpha}\overline{\bf x}{\bf y}|\\
 <{\bf \mu}\hspace{2mm}|&=&
 <\overline{\bf \alpha}\overline{\bf x}
 \overline{\bf x}{\bf y}|,
\end{array}
\right.
\hspace{0.7cm}\mbox{2nd generation}\\
&&\left\{
\begin{array}{lcl}
 <{\bf t}|&=&<{\bf \alpha}{\bf x}{\bf y}{\bf y}|\\
 <{\bf b}|&=&<\overline{\bf \alpha}\overline{\bf x}
 {\bf x}{\bf y}{\bf y}|,
\end{array}
\right.
\hspace{0.3cm}
\left\{
\begin{array}{lcl}
 <{\bf \nu_\tau}|&=&
 <\overline{\bf \alpha}\overline{\bf x}{\bf y}{\bf y}|\\
 <{\bf \tau}\hspace{2mm}|&=&
 <\overline{\bf \alpha}\overline{\bf x}
 \overline{\bf x}{\bf y}{\bf y}|,
\end{array}
\right.
\hspace{4mm}\mbox{3rd generation},\label{11}
\end{eqnarray} 
\setcounter{equation}{\value{enumi}}}
 where the suffix $L,R$s are omitted for brevity. 
 We can also make vector and scalar particles 
 with (1,1,1) : 
{
\setcounter{enumi}{\value{equation}}
\addtocounter{enumi}{1} 
\setcounter{equation}{0}
\renewcommand{\theequation}{\theenumi\alph{equation}}
\begin{eqnarray}&&\left\{
\begin{array}{lcl}
 <{\bf W}^+|&=&
 <{\bf \alpha}^\uparrow{\bf \alpha}^\uparrow{\bf x}|\\
 <{\bf W}^-|&=&
 <\overline{\bf \alpha}^\uparrow\overline{\bf \alpha}
 ^\uparrow\overline{\bf x}|,
\end {array}
\right.\hspace{6mm}
\left\{
\begin{array}{lcl}
 <{\bf Z}_1^0|&=&
 <{\bf \alpha}^\uparrow\overline{\bf \alpha}^\uparrow\\
 <{\bf Z}_2^0|&=&
 <{\bf \alpha}^\uparrow\overline{\bf \alpha}^\uparrow
 {\bf x}\overline{\bf x}|, 
\end{array} 
\right.\hspace{1cm}\mbox{Vector}\\
&&\left\{
\begin{array}{lcl}
 <{\bf S}^+|&=&
 <{\bf \alpha}^\uparrow{\bf \alpha}^\downarrow{\bf x}|\\
 <{\bf S}^-|&=&
 <\overline{\bf \alpha}^\uparrow\overline{\bf \alpha}
 ^\downarrow{\bf x}|,
\end{array}
\right.
\hspace{9mm}
\left\{
\begin{array}{lcl}
 <{\bf S}_1^0|&=&
 <{\bf \alpha}^\uparrow\overline{\bf \alpha}^\downarrow|\\
 <{\bf s}_2^0|&=&
 <{\bf \alpha}^\uparrow\overline{\bf \alpha}^\downarrow
 {\bf x}\overline{\bf x}|,
\end{array}
\right.
\hspace{8mm}\mbox{Scalar},\label{12}
\end{eqnarray}
\setcounter{equation}{\value{enumi}}}
 where the suffix $L,R$s are omitted for brevity and 
 $\uparrow, \downarrow$ indicate $spin\hspace{1mm}up, 
 spin\hspace{1mm}down$ states.
 They play the role of intermediate bosons same as 
 ${\bf \pi}$, ${\bf \rho}$ in the strong interactions. 
 As Eq.(9) and Eq.(12) contain only ${\bf \alpha}$ and 
 ${\bf x}$ subquarks, we can draw the 
 ``$line\hspace{1mm}diagram$''s of weak interactions 
 as seen in Fig (1). 
 Eq.(9,d) shows that the electron is constructed 
 from antimatters only. We know, phenomenologically, 
 that this universe is mainly made of protons, electrons, 
 neutrinos, antineutrinos and unknown dark matters. 
 It is said that protons and electrons in the universe are 
 almost same in quantity. 
 Our model show that one proton has the configuration
 of $({\bf u}{\bf u}{\bf d})=(2{\bf \alpha}, 
 \overline{\bf \alpha}, 3{\bf x}, \overline{\bf x})$; 
 electron : $(\overline{\bf \alpha}, 2\overline{\bf x})$; 
 neutrino : $({\bf \alpha}, \overline{\bf x})$; 
 antineutrino : $(\overline{\bf \alpha}, {\bf x})$ and 
 the dark matters are presumably constructed from the same 
 amount of matters and antimatters because of their neutral 
 charges. 
 Therefore these facts may lead the thought that 
 ``the universe is the matter-antimatter-even object.'' 
 And then there exists a conception-leap between 
 ``proton-electron abundance'' and ``matter abundance'' 
 if our composite scenario is admitted (as for the possible 
 way to realize the proton-electron excess universe, 
 see Ref.[5]).
\par
 Our composite model contains two steps, 
 namely the first is ``subquarks made of primons'' 
 and the second is ``quarks and leptons made of subquarks''. 
 Here let us discuss about the mass generation mechanism  
 of quarks and leptons as composite objects. 
 Our model has only one  kind of fermion : $\Lambda$  
 and boson : $\Theta$. 
 The first step of ``subquarks made of primons'' seems 
 to have nothing to do with 'tHooft's anomaly matching 
 condition[22] because there is no global symmetry with 
 $\Lambda$ and $\Theta$. Therefore from this line of 
 thought it is  impossible to say anything about that 
 ${\bf \alpha}$, ${\bf \beta}$, ${\bf x}$ and ${\bf y}$ 
 are massless or massive. 
 However, if it is the case that the neutral  
 (1,1,1)-states of primon-antiprimon composites 
 (as is stated above) become the dark matters, 
 the masses of them presumably be less than the order of 
 MeV from the phenomenological aspects of astrophysics. 
 Then we may assume that these subquarks are massless 
 or almost massless compared with $\Lambda_{L,R}$ 
 in practice, that is, utmost a few MeV. 
 In the second step, the arguments of 'tHooft's anomaly 
 matching condition are meaningful. 
 The confining of subquarks must occur at the energy scale 
 of $\Lambda_{L,R}>>\Lambda_c$ and then it is natural that 
 $\alpha_s, \alpha \rightarrow0$ and that the gauge 
 symmetry group is the spontaneously unbroken 
 $SU(2)_L\otimes{SU(2)}_R$ gauge group. 
 Seeing Eq.(9), we find quarks and leptons are composed of 
 the mixtures of subquarks and antisubquarks. 
 Therefore it is proper to regard subquarks and 
 antisubquarks as different kinds of particles. 
 From Eq.(8,a,b) we find eight kinds of fermionic 
 subquarks ( 3 for ${\bf \alpha}$, 
 $\overline{\bf \alpha}$ and 1 for ${\bf \beta}$, 
 $\overline{\bf \beta}$). So the global 
 symmetry concerned is $SU(8)_L\otimes{SU(8)}_R$. 
 Then we arrange :
\begin{equation}
 ({\bf \beta},\overline{\bf \beta},{\bf \alpha}_i,
 \overline{\bf \alpha}_i\hspace{3mm}i=1,2,3
 \hspace{1mm})_{L,R}\hspace{2cm}in
 \hspace{1cm}(SU(8)_L\otimes{SU(8)}_R)_{global},\label{13}
\end{equation}
 where $i$s are color freedoms.
 Next, the fermions in Eq.(12) are confined 
 into the singlet states of the local $SU(2)_L
 \otimes{SU(2)}_R$ gauge symmetry and make up quarks and 
 leptons as seen in Eq.(9) (eight fermions). 
 Then we arrange :
\begin{equation}
 ({\bf \nu_e},{\bf e},{\bf u}_i,{\bf d}_i\hspace{3mm}
 i=1,2,3\hspace{1mm})_{L,R}\hspace{2cm}in\hspace{1cm}
 (SU(8)_L\otimes{SU(8)}_R)_{global},\label{14}
\end{equation}
 where $i$s are color freedoms. From Eq.(13) and Eq.(14) 
 the anomalies of the subquark level and the quark-lepton 
 level are matched and then all composite quarks and 
 leptons (in the 1st generation) are remained massless. 
 Note again that presumably, ${\bf \beta}$ and 
 $\overline{\bf \beta}$ in Eq.(12) are composed 
 into ``bosonic'' (${\bf \beta}$${\bf \beta}$), 
 (${\bf \beta}$$\overline{\bf \beta}$) and 
 ($\overline{\bf \beta}$$\overline{\bf \beta}$), 
 which vapor out to the dark matters. 
 Schrempp and Schrempp have discussed about a confining 
 $SU(2)_L\otimes{SU(2)}_R$ gauge model with three 
 fermionic preons and stated that it is possible that 
 not only the left-handed quarks and leptons are 
 composite but also the right-handed are so on the 
 condition that $\Lambda_R/\Lambda_L$ is at least of the 
 order of $3$[17].
 If CDF's data[23] truly indicates the compositeness of 
 quarks, $\Lambda_L$ is presumably around $1.6$ TeV. 
 As seen in Eq.(12.a) the existence of composite 
 ${\bf W}_R$, ${\bf Z}_R$ is predicted. 
 As concerning, the fact that they are not observed yet 
 means that the masses of ${\bf W}_R$, ${\bf Z}_R$ are 
 larger than those of ${\bf W}_L$, ${\bf Z}_L$ and that 
 $\Lambda_R>\Lambda_L$. 
 Owing to 'tHooft's anomaly matching condition the small 
 mass nature of the 1st generation comparing to $\Lambda_L$ 
 is guaranteed but the evidence that the quark masses of 
 the 2nd and the 3rd generations become larger as the 
 generation numbers increase seems to have nothing to do 
 with the anomaly matching mechanism in our model, 
 because as seen in Eq.(11,a,b) these generations are 
 obtained by just adding scalar ${\bf y}$-particles. 
 This is different from Abott and Farhi's model in which 
 all fermions of three generations are equally embedded 
 in $SU(12)$ global symmetry group and all members take 
 part in the anomaly matching mechanism[15,25]. 
 Concerning this, let us discuss a little about subquark 
 dynamics inside quarks. 
 According to ``Uncertainty Principle'' the radius of 
 the composite particle is, in general, roughly inverse 
 proportional to the kinetic energy of the constituent 
 particles moving inside it. 
 The radii of quarks may be around $1/\Lambda_{L,R}$ .
 So the kinetic energies of subquarks may be more than 
 hundreds GeV and then it is considered that the masses of 
 quarks essentially depend on the kinetic energies of 
 subquarks and such a large binding energy as 
 counterbalances them. As seen in Eq.(10,a,b) our model 
 shows that the more the generation number increases 
 the more the number of the constituent particles increases. 
 So assuming that the radii of all quarks do not vary so 
 much (because we have no experimental evidences yet), 
 the interaction length among subquarks inside quarks 
 becomes shorter as generation numbers increase and 
 accordingly the average kinetic energy per one 
 subquark may increase. 
 Therefore integrating out the details of subquark dynamics 
 it cloud be said that the essential feature of increasing 
 masses of the 2nd and the 3rd generations is simply because 
 their masses are described as a increasing function of 
 the sum of the kinetic energies of constituent subquarks. 
 From the Review of Particles and Fields[29] 
 we can phenomenologically parameterized the mass 
 spectrum of quarks and leptons as follows : 
{
\setcounter{enumi}{\value{equation}}
\addtocounter{enumi}{1}
\setcounter{equation}{0}
\renewcommand{\theequation}{\theenumi\alph{equation}}
\begin{eqnarray}
 M_{UQ}&=&
 1.2\times10^{-4}\times(10^{2.05})^n\hspace{1cm}\mbox{GeV}
 \hspace{1.5cm}\mbox{for}\hspace{2mm}\mbox{{\bf u},
 {\bf c},{\bf t}},\\
 M_{DQ}&=&
 3.0\times10^{-4}\times(10^{1.39})^n\hspace{1cm}\mbox{GeV}
 \hspace{1.5cm}\mbox{for}\hspace{2mm}\mbox{{\bf d},
 {\bf s},{\bf b}},\\
 M_{DL}&=&
 3.6\times10^{-4}\times(10^{1.23})^n\hspace{1cm}\mbox{GeV}
 \hspace{1.5cm}\mbox{for}\hspace{2mm}\mbox{{\bf e},
 ${\bf \mu}$,${\bf \tau}$},\label{15}
\end{eqnarray}
\setcounter{equation}{\value{enumi}}}
 where $n=1,2,3$ are the generation numbers. 
 They seem to be geometric-ratio-like. 
 The slopes of the up-quark sector and down-quark sector 
 are different, so it seems that each has different 
 aspects in subquark dynamics. 
 From Eq.(15) we obtain $M_{\bf u}=13.6$ MeV, 
 $M_{\bf d}=7.36$ MeV and $M_{\bf e}=6.15$ MeV. 
 These are a little unrealistic compared with the 
 experiments[4]. 
 But considering the above discussions about the anomaly 
 matching conditions (Eq.(13,14)), it is natural that the 
 masses of the members of the 1st generation are roughly 
 equal to those of the subquarks, that is, a few MeV. 
 The details of their mass-values depend on the subquark 
 dynamics owing to the effects of electromagnetic and 
 color gauge interactions. These mechanism has studied by 
 Weinberg[26] and Fritzsch[27].
\par
 One of the experimental evidences inspiring the SM is 
 the ``universality'' of the coupling strength among the 
 weak interactions. Of course if the intermediate bosons 
 are gauge fields, they couple to the matter fields 
 universally. 
 But the inverse of this statement is not always true, 
 namely the quantitative equality of the coupling 
 strength of the interactions does not necessarily imply 
 that the intermediate bosons are elementary gauge bosons. 
 In practice the interactions of ${\bf \rho}$ and 
 ${\bf \omega}$ are regarded as indirect 
 manifestations of QCD. 
 In case of chiral $SU(2)\otimes{SU(2)}$ the pole dominance 
 works very well and the predictions of current algebra and 
 PCAC seem to be fulfilled within about $5$\%[18]. 
 Fritzsch and Mandelbaum[16,18] and Gounaris, K\"ogerler 
 and Schildknecht[28,29] have elaborately discussed about 
 universality of weak interactions appearing as a 
 consequence of current algebra and ${\bf W}$-pole 
 dominance of the weak spectral functions from the stand 
 point of the composite model. 
 Extracting the essential points from their  
 arguments we shall mention the followings . 
\par
 In the first generation let the weak charged currents be 
 written in terms of the subquark fields as :
\begin{equation}
 {\bf J}_{\mu}^{+}=\overline{U}h_{\mu}D,\hspace{2cm}
 {\bf J}_{\mu}^{-}=\overline{D}h_{\mu}U,\label{16}
\end{equation}
 where $U=({\bf \alpha}{\bf x})$, 
 $D=(\overline{\bf \alpha}\overline{\bf x}{\bf x})$ 
 and $h_{\mu}=\gamma_{\mu}(1-\gamma_5)$.
 Reasonableness of Eq.(16) may given by the fact that
 $M_W<<\Lambda_{L,R}$ (where $M_W$ is 
 ${\bf W}$-boson mass).
 Further, let $U$ and $D$ belong to the doublet of 
 the global weak isospin $SU(2)$ group and ${\bf W}^+$, 
 ${\bf W}^-$, $1/\sqrt{2}({\bf Z}_1^0-{\bf Z}_2^0)$ be 
 in the triplet and $1/\sqrt{2}({\bf Z}_1^0+{\bf Z}_2^0)$ 
 be in the singlet of $SU(2)$. These descriptions seem to 
 be natural if we refer the diagrams in Fig.(1). 
 The universality of the weak interactions are inherited 
 from the universal coupling strength of the algebra of 
 the global weak isospin $SU(2)$ group with the assumption 
 of ${\bf W}$-, ${\bf Z}$-pole dominance. 
 The universality including the 2nd and the 3rd 
 generations are based on the above assumptions and the 
 concept of the flavor-mixings.
 The quark-flavor-mixings in the weak interactions are 
 expressed by Cabbibo-Kobayashi-Maskawa (CKM)-matrix 
 ($V_{ij}$) based on the SM.  
 Its nine matrix elements (in case of three generations) 
 are free parameters (in practice four parameters with 
 the unitarity) and this point is said to be one of the 
 drawback of the SM along with non-understanding of 
 the origins of the quark-lepton mass spectrum and 
 generations. 
 In the SM, the quark fields (lepton fields also) are 
 elementary and then we are able to investigate, 
 at the utmost, the external relationship among them. 
 On the other hand if quarks  are the composites of 
 substructure constituents, the quark-flavor-mixing 
 phenomena must be understood by the substructure 
 dynamics and the values of CKM matrix elements become 
 materials for studying these . 
 In our model ``\begin{em}the quark-flavor-mixings occur 
 by creations or annihilations of ${\bf y}$-particles 
 inside quarks\end{em}''. 
 The ${\bf y}$-particle is a neutral scalar subquark in 
 the {\bf 3}-state of $SU(2)_L$ group and then couples to 
 two hypercolor gluons (denoted by ${\bf g}_h$). 
 By this mechanism we obtained$|V_{ub}|=3.45\times10^{-3}$,
 $|V_{ts}|=2.62\times10^{-2}$ and $|V_{td}|=1.40
 \times10^{-3}$(for detailed analysis see Ref.[52]). 
 
\section{Mass differences and CP-Violations 
 in Neutral Meson Systems}
\hspace{\parindent}
\par
 {\bf a. Mass difference ${\Delta}M_P$
 by $P^{0}-\overline{P^{0}}$ mixing} 
\newline
\par
 The typical ${\Delta}F=2$ phenomenon is the mixing 
 between a neutral pseudo scalar meson ($P^0$) and its 
 antimeson ($\overline{P^{0}}$). There are six types of 
 them, e.g., $K^{0}-\overline{K^{0}}$, 
 $D^{0}-\overline{D^{0}}$, $B^{0}_{d}-\overline{B^{0}_{d}}$, 
 $B^{0}_{s}-\overline{B^{0}_{s}}$, $T^{0}_{u}-
 \overline{T^{0}_{u}}$ and $T^{0}_{c}-\overline{T^{0}_{c}}$ 
 mixings. 
 Usually they have been considered to be the most sensitive 
 probes of higher-order effects of the weak interactions 
 in the SM. 
 The basic tool to investigate them is the ``box diagram''. 
 By using this diagram to the $K_{L}$-$K_{S}$ mass 
 difference, Gaillard and Lee predicted the mass of the 
 charm quark[30]. Later, Wolfenstein suggested that the 
 contribution of the box diagram which is called the 
 short-distance (SD) contribution cannot supply the whole 
 of the mass difference ${\Delta}M_K$ and there are 
 significant contributions arising from the long-distance 
 (LD) contributions associated with low-energy 
 intermediate hadronic states[31]. 
 As concerns, the LD-phenomena occur in the energy range 
 of few hundred MeV and the SD-phenomena around $100$ GeV 
 region. 
 Historically there are various investigations for 
 $P^{0}$-$\overline{P^{0}}$ mixing problems[3][32-41] and 
 many authors have examined them by use of LD- and 
 SD-contributions. 
 In summary, the comparison between the theoretical results 
 and the experiments about ${\Delta}M_P$ 
 ($P=K,D$ and $B_d$) are as follows :
{  
\setcounter{enumi}{\value{equation}}
\addtocounter{enumi}{1}
\setcounter{equation}{0}
\renewcommand{\theequation}{\theenumi\alph{equation}}
\begin{eqnarray}
 {\Delta}M^{LD}_{K}&\approx&{\Delta}M^{SD}_{K}\approx
 {\Delta}M^{exp}_{K},\\
 {\Delta}M^{SD}_{D}&\ll&{\Delta}M^{LD}_{D}
 (\ll{\Delta}M^{exp}_{D},upper\hspace{2mm}bound),\\
 {\Delta}M^{LD}_{B_d}&\ll&{\Delta}M^{SD}_{B_d}\simeq
 {\Delta}M^{exp}_{B_d}.\label{17}
\end{eqnarray}
\setcounter{equation}{\value{enumi}}}
 Concerning Eq.(17a) it is explain that ${\Delta}M_{K}=
 {\Delta}M^{SD}_{K}+D{\Delta}M^{LD}_{K}$ where ``$D$'' is 
 a numerical value of order $O(1)$. As for Eq(17c), 
 they found that ${\Delta}M^{LD}_{B_d}\approx10^{-16}$ GeV 
 and ${\Delta}M^{SD}_{B_d}\approx10^{-13}$ GeV, 
 then the box diagram is the most important 
 for $B^{0}_{d}$-$\overline{B^{0}_{d}}$ mixing. 
 Computations of ${\Delta}M^{SD}_{B_d}$ and 
 ${\Delta}M^{SD}_{B_s}$ from the box diagrams 
 in the SM give
\begin{equation}
 {\Delta}M^{SD}_{B_s}/{\Delta}M^{SD}_{B_d}\simeq
 (M_{B_s}/M_{B_d})|V_{ts}/V_{td}|^2
 (B_{B_s}{f}^2_{B_s}/B_{B_d}{f}^2_{B_d})\zeta,
 \label{18}
\end{equation}
 where $V_{ij}$s stand for CKM matrix elements; $M_P$ : 
 P-meson mass; $\zeta$ : a QCD correction of order $O(1)$; 
 $B_B$ : Bag factor of B-meson and $f_B$ : decay constant 
 of B-meson.
 Measurements of ${\Delta}M^{exp}_{B_d}$ and 
 ${\Delta}M^{exp}_{B_s}$ are, therefore, said to be useful 
 to determine $|V_{ts}/V_{td}|$[42][43]. 
 Concerning Eq.(17b), they found that ${\Delta}M^{LD}_{D}
 \approx10^{-15}$ GeV and ${\Delta}M^{SD}_{D}
 \approx10^{-17}$ GeV[3][37] but the experimental 
 measurement is ${\Delta}M^{exp}_{D}<1.3\times10^{-13}$ 
 GeV[4]. 
 Further there is also a study that ${\Delta}M^{LD}_D$ is 
 smaller than $10^{-15}$ GeV by using the heavy quark 
 effective theory[38]. 
 Then many people state that it would be a signal of new 
 physics beyond the SM if the future experiments confirm 
 that ${\Delta}M^{exp}_D\simeq10^{-14}\sim10^{-13}$ 
 GeV[32-38].
 Above investigations are based on the calculations of 
 SD-contributions with (or without) LD-contributions 
 in the SM. 
\par
 On the other hand some authors have studied these 
 phenomena in the context of the theory explained by 
 the single dynamical origin. 
 Cheng and Sher[44], Liu and Wolfenstein[40], and G\'erard 
 and Nakada[41] have thought that all 
 $P^{0}$-$\overline{P^0}$ mixings occur only by the 
 dynamics of the TeV energy region which is essentially 
 the same as the super-weak (SW) idea originated by 
 Wolfenstein[2]. 
 They extended the original SW-theory (which explains 
 CP violation in the $K$-meson system) to other flavors by 
 setting the assumption that ${\Delta}F=2$ changing 
 neutral $spin\hspace{2mm}0$ particle with a few TeV mass 
 (denoted by $H$) contributes to the ``real part'' of 
 $M_{ij}$ which determines ${\Delta}M_P$ and also the 
 ``imaginary part'' of $M_{ij}$ which causes the indirect 
 CP violation. 
 The ways of extensions are that $H$-particles couple to 
 quarks by the coupling proportional to $\sqrt{{m}_i
 {m}_j}$[40][44], $({m}_i/{m}_j)^n\hspace
 {2mm}n=0,1,2$[40] and $({m}_i+{m}_j)$[41] 
 where $i,j$ are flavors of quarks coupling to $H$. 
 It is suggestive that the SW-couplings depend on quark 
 masses (this idea is adopted in our model discussed 
 below). Cheng and Sher[44] and Liu and Wolfenstein[40] 
 obtained that ${\Delta}M_D=({m}_c/{m}_s)
 {\Delta}M^{exp}_{K}\approx10^{-14}$ GeV with the 
 assumption that $H$-exchange mechanism saturates the 
 ${\Delta}M^{exp}_K$ bound, which is comparable to 
 ${\Delta}M^{exp}_{D}<1.3\times10^{-13}$ GeV[4].
 Concerning $B$-meson systems they found that 
 ${\Delta}M_{B_s}/{\Delta}M_{B_d}={m}_s/{m}_d
 \simeq20$ which seems agreeable to $({\Delta}M_{B_s}/
 {\Delta}M_{B_d})_{exp}>13$[43]. 
 However using their scheme it is calculated that
\begin{equation}
 {\Delta}M_{B_d}/{\Delta}M_K=(B_{B_d}{f}^2_{B_d}/
 B_K{f}^2_K)(M_{B_d}/M_K)(m_b/m_s)
 \simeq300,\label{19}
\end{equation}
 where we use $m_b=4.3$ GeV, $m_s=0.2$ GeV, 
 $M_{B_d}=5.279$ GeV, $M_K=0.498$ GeV, 
 $B_{B_d}{f}^2_{B_d}=(0.22{\rm GeV})^2$, 
 $B_K{f}^2_K=(0.17{\rm GeV})^2$. 
 It seems larger than 
 $({\Delta}M_{B_d}/{\Delta}M_K)_{exp}=85.8\pm3.6$[43]. 
 This result is caused by rather large b-quark mass value. 
\par
 Now let us discuss $P^0$-$\overline{P^0}$ mixings by 
 using our FB-model. The discussions start from the 
 assumption that the mass mixing matrix $M_{ij}(P)$ $(i(j)=
 1(2)$ denotes $P^0(\overline{P^0}))$ is saturated by the 
 super-weak-type interactions causing a direct 
 ${\Delta}F=2$ transitions. 
 We usually calculate ${\Delta}M_P$ as 
{  
\setcounter{enumi}{\value{equation}}
\addtocounter{enumi}{1}
\setcounter{equation}{0}
\renewcommand{\theequation}{\theenumi\alph{equation}}
\begin{eqnarray}
 M_{12}(P)&=
 &<\overline{P^0}|{\cal H}^{{\Delta}F=2}_{SW}|P^0>,\\
 {\Delta}M_P\hspace{2mm}&=
 &M_H-M_L\simeq2|M_{12}(P)|,\label{20}
\end{eqnarray}
\setcounter{equation}{\value{enumi}}}
 where we assume $ImM_{12}\ll{ReM_{12}}$ which is 
 experimentally preferable[3][45], and $M_{H(L)}$ stands 
 for heavier (lighter) $P^0$-meson mass. 
 Applying the vacuum-insertion calculation to the hadronic 
 matrix element as $<\overline{P^0}|[\overline{{q}_{i}}
 \gamma_{\mu}(1-\gamma_5){q}_j]^2|P^0>\sim
 B_P{f}^2_{P}M^2_P$[3] we get
\begin{equation}
 M_{12}(P)=(1/12\pi^2)B_P{f}^2_{P}M_{P}{\cal M}_P.
 \label{21}
\end{equation}
 The details of ${\cal M}_P$ are model-dependent, e.g., 
 the box diagram in the SM; the neutral $spin\hspace{2mm}0$ 
 particle exchange in the SW-theory. 
 In case of our FB-model, the diagrams contributing to 
 ${\cal M}_P$ are seen in Fig.(2). In our model 
 $P^0$-$\overline{P^0}$ mixings occur due to 
 ``{\bf y}-exchange'' between two quarks inside the present 
 $P^0$-meson. 
 This is a kind of the realizations of Wolfenstein's 
 SW-idea[2]. The schematic illustration is as follows : 
 two particles (quarks) with radius order of 
 $1/{\Lambda}_q$ (a few ${\rm TeV}^{-1}$) are moving to and 
 fro inside a sphere (meson) with radius order of 
 ${\rm GeV}^{-1}$. 
 The ${\bf y}$-exchange interactions would occur when 
 two quarks inside $P^0$-meson interact in contact with 
 each other because ${\bf y}$-particles are confined 
 inside quarks. 
 As seen in Fig.(2), the contributions of 
 ${\bf y}$-exchanges seem common among various $P^0$-mesons. 
 Upon this, setting the assumption :
 ``\begin{em}universality of the ${\bf y}$-exchange
 interactions\end{em}'', we rewrite ${\cal M}_P$ as
\begin{equation}
 {\cal M}_P={n}_P{\eta}(P)
 {\tilde {\cal M}}_{i}(P),\label{22}
\end{equation}
 where $n_P=1$ for $P=K, D, B_d, T_u$; $n_P=2$ for $P=B_s, 
 T_c$, $i=1$ for $K, D, B_s, T_u$; $i=2$ for $B_d, T_c$.
 Then the universality means explicitly that 
{  
\setcounter{enumi}{\value{equation}}
\addtocounter{enumi}{1}
\setcounter{equation}{0}
\renewcommand{\theequation}{\theenumi\alph{equation}}
\begin{eqnarray}
 {\tilde {\cal M}}_{1}(K)&=&{\tilde {\cal M}}_{1}(D)=
 {\tilde {\cal M}}_{1}(B_s)={\tilde {\cal M}}_{1}(T_c),\\
 {\tilde {\cal M}}_{2}(B_d)&=&{\tilde {\cal M}}_{2}(T_u).
 \label{23}
\end{eqnarray}
\setcounter{equation}{\value{enumi}}}
 The explanation of ${n}_P$ is such that $K$ and $D$ 
 have one ${\bf y}$-particle and one ${\bf y}$-particle 
 exchanges; $B_d$ and $T_u$ have two ${\bf y}$-particles 
 and both of them exchange simultaneously, so for them 
 we set ${n}_P=1$ and $B_s$ and $T_c$ have two 
 ${\bf y}$-particles but one of them exchanges, 
 so they have ${n}_P=2$ because the probability becomes 
 double. 
 The `` ${i}$ '' means the number of exchanging 
 ${\bf y}$-particles in the present diagram. 
 Concerning ${\eta}(P)$, we shall explain as follows : 
 In our FB-model $P^0$-$\overline{P^0}$ mixing occurs by 
 the ``contact interaction'' of two quarks coliding 
 inside $P^0$-meson. 
 Therefore the probability of this interaction may be 
 considered inverse proportional to the volume of the 
 present $P^0$-meson, e.g., the larger radius $K$-meson 
 gains the less-valued probability of the coliding than 
 the smaller radius $D$- (or $B_s$-) meson. 
 The various aspects of hadron dynamics seem to be 
 successfully illustrated by the semi-relativistic 
 picture with ``Breit-Fermi Hamiltonian''[46]. 
 Assuming the power-law potential 
 $V(r)\sim{r}^{\nu}$($\nu$ is a real number), the radius 
 of $P^0$-meson (denoted by ${\bf r}_P)$ is  
 proportional to ${\mu}_P^{-1/(2+\nu)}$, where 
 ${\mu}_P$ is the reduced mass of two quark-masses inside 
 $P^0$-meson[46]. Then the volume of $P^0$-meson is 
 proportional to ${\bf r}_P^{3}\sim{\mu}_P^{-3/(2+\nu)}$. 
 After all we could assume for ${\eta}(P)$ in Eq.(22) as
{  
\setcounter{enumi}{\value{equation}}
\addtocounter{enumi}{1}
\setcounter{equation}{0}
\renewcommand{\theequation}{\theenumi\alph{equation}}
\begin{eqnarray}
 {\eta}(P)&=&{\xi}({\mu}_P/
 {\mu}_K)^{1}\hspace{2cm}{\rm for\hspace{3mm}
 linear-potential},\\
 &=&{\xi}({\mu}_P/
 {\mu}_K)^{1.5}\hspace{1.7cm}{\rm for\hspace{8mm}
 log-potential},\label{24}
\end{eqnarray}
\setcounter{equation}{\value{enumi}}}
 where $\xi$ is a dimensionless numerical factor 
 depending on the details of the dynamics and 
 ${\eta}(P)$ is normalized by ${\mu}_K$ (reduced mass 
 of $s$- and $d$-quark in $K$ meson) for convenience. 
 We may think that the ${\bf y}$-exchange  
 is described by the overlapping of the wave functions of 
 two quarks inside $P^0$-meson. Then we shall write as 
{  
\setcounter{enumi}{\value{equation}}
\addtocounter{enumi}{1}
\setcounter{equation}{0}
\renewcommand{\theequation}{\theenumi\alph{equation}}
\begin{eqnarray}
 |\tilde{{\cal M}}_i (P)|
        &=&
(1/\Lambda^2_q)
| \kappa \int \Psi_q(r) \Psi_{q^\prime}(r) d^3 r |,
\\
        &\simeq&
|\kappa \int \Psi_q(r) \Psi_{q^\prime}(r) d^3 r |
 \times 10^{-7} \hspace{1cm} {\rm GeV}^{-2},
\label{25}
\end{eqnarray}
\setcounter{equation}{\value{enumi}}}
 where $\Psi_q (r)$ is a radial wave function of 
 $q$-quark, $\kappa$ is a dimensionless complex numerical 
 factor caused by unknown subquark dynamics and may depend 
 on $|<q^{'}|{\overline{\bf y}}(\partial_{\mu}{\bf y})|q>|$.
 In Eq.(25b) we estimate a few TeV as $\Lambda_q$. 
\par 
 From the experimental informations the complex 
 $M_{12}(K)$ is evaluated as [3] 
\begin{equation} 
 M_{12}^{exp}(K)=
   -(0.176+i0.114\times10^{-2})\times10^{-14}
   \hspace{1.5cm}{\rm Gev}.\label{26}
\end{equation}
 On the other hand, setting $P$=$K$ in Eq.(21) we obtain 
\begin{equation}
 |{\cal M}_K|=
    |M_{12}(K)|/(1/12 \pi^2)B_K{f}^2_KM_K  
             \simeq0.15\times10^{-10} 
             \hspace{1.5cm}{\rm GeV}^{-2},\label{27}
\end{equation}
 where we use $|M_{12}(K)|=|M_{12}^{exp}(K)|$ 
 from Eq.(26), $B_K{f}_K^2=(0.17{\rm GeV})^2$ and 
 $M_K=0.498$ GeV. 
 Further setting $P=K$ in Eq.(22), (24), and (25) 
 we have 
\begin{equation} 
|{\cal M}_K|={\xi}|\tilde{{\cal M}}_{1}(K)|
        =
 {\xi}|\kappa \int \Psi_s(r) \Psi_d(r) d^3 r |
 \times 10^{-7} \hspace{1cm} {\rm GeV}^{-2},
 \label{28}
\end{equation}
 From Eq.(27) and (28) we obtain 
\begin{equation}
  {\xi}|\kappa \int \Psi_s(r) \Psi_d(r) d^3 r |
 \simeq10^{-4}\hspace{1cm} {\rm GeV}^{-2}.\label{29}
\end{equation}
 As it may be expected that 
\begin{equation}
  | \int \Psi_s(r) \Psi_d(r) d^3 r |
     \simeq O(10^{-2}) \sim O(10^{-1}),\label{30}
\end{equation}
 we have 
\begin{equation}
  {\xi}|\kappa|\simeq O(10^{-3}) \sim O(10^{-2}).
 \label{31} 
\end{equation}
 Eq.(30) has to be ascertained in future. 
 If the above investigation is the case, our 
 picture corresponds to the scheme of the Higgs 
 (of a few TeV mass value) exchange in 
 the SW-theory[40][44][47][49][50]. 
 Note that the mass value of ${\bf y}$-particle itself 
 is less than a few MeV as seen in Sect.(3). 
\par 
 The experimental results of $\Delta M_P$ are as 
 follows[4][43] : 
{  
\setcounter{enumi}{\value{equation}}
\addtocounter{enumi}{1}
\setcounter{equation}{0}
\renewcommand{\theequation}{\theenumi\alph{equation}}
\begin{eqnarray}
 {\Delta}M_{K}&=&(3.510\pm0.018) \times 10^{-15}
                 \hspace{2cm}{\rm GeV},\\
 {\Delta}M_{D}&<&1.3 \times 10^{-13}
                 \hspace{4.2cm}{\rm GeV},\\
 {\Delta}M_{B_d}&=&(3.01 \pm 0.13) \times 10^{-13}
                 \hspace{2.4cm}{\rm GeV},\\
{\Delta}M_{B_s}&>&4.0 \times 10^{-12}
                 \hspace{4.2cm}{\rm GeV}.\label{32}
\end{eqnarray}
\setcounter{equation}{\value{enumi}}}
 Using Eq.(20), (21) and (32), we have
{  
\setcounter{enumi}{\value{equation}}
\addtocounter{enumi}{1}
\setcounter{equation}{0}
\renewcommand{\theequation}{\theenumi\alph{equation}}
\begin{eqnarray}
 |{\cal M}_D|     &<& \hspace{2mm}8.1  |{\cal M}_K|,\\
 |{\cal M}_{B_d}| &=&            4.86 |{\cal M}_K|,\\
 |{\cal M}_{B_s}| &>&            49.3 |{\cal M}_K|.
 \label{33}
\end{eqnarray}
\setcounter{equation}{\value{enumi}}}
 At the level of ${\cal M}_P$, it seems that 
\begin{equation} 
  |{\cal M}_P|/|{\cal M}_K| \simeq O(1) \sim O(10),
 \label{34}
\end{equation}
 where $P=D, B_d, B_s$. 
\par 
 Here let us go on to more precise investigations. 
 In Eq.(23) let us extend the ``universality'' and 
 assume that
 $|{\tilde {\cal M}}_1(K)| \simeq |
 {\tilde {\cal M}}_2(B_d)|$.
 Using Eq.(22), (24) and (33b) we obtain 
{  
\setcounter{enumi}{\value{equation}}
\addtocounter{enumi}{1}
\setcounter{equation}{0}
\renewcommand{\theequation}{\theenumi\alph{equation}}
\begin{eqnarray}
 {\mu}_{B_d}/{\mu}_K   &=&  4.86
                \hspace{3cm}{\rm for\hspace{3mm}
 linear-potential},\\
                                     &=&  2.87
                \hspace{3cm}{\rm for\hspace{8mm}
 log-potential},\label{35}           
\end{eqnarray}
\setcounter{equation}{\value{enumi}}}
 where $B_{B_d}{f}^2_{B_d}=(0.22{\rm GeV})^2$, 
 $B_K{f}^2_K=(0.17{\rm GeV})^2$ are used. 
 Note that, comparing with the case of Eq.(19), 
 we can evade the large enhancement by $b$-quark mass 
 effect. 
 This is because the quark mass dependence is 
 introduced through the reduced mass (in which the 
 effect of heavier mass decreases). 
 Some discussions are as follows : 
 If we adopt the pure non-relativistic picture 
 it may be that 
 ${\mu}_K \simeq {\mu}_{B_d} \simeq {m}_d
 \simeq ({\mu}_D \simeq {\mu}_{T_u})$ but from the 
 semi-relativistic standpoint it seems preferable that 
 ${\mu}_K (< {\mu}_{D}) < 
  {\mu}_{B_d}(<{\mu}_{T_u})$ because the effective 
 mass value of ``$d$-quark'' in $B_d$-meson is considered 
 larger than that in $K$-meson. 
 It may be caused by that the kinetic energy of 
 ``$d$-quark'' in $B_d$-meson is larger than that in 
 $K$-meson owing to the fact : ${\bf r}_{B_d}<{\bf r}_K$. 
 Then we can expect the plausibility of Eq.(35). 
 Of course it may be also a question whether the extension 
 of the universality : 
 $|{\tilde {\cal M}}_{1}(K)|\simeq
 |{\tilde {\cal M}}_{2}(B_d)|$
 is good or not (this point may influence Eq.(35)),
 which will become clear when the experimental 
 result about ${\Delta}M_{T_u}$ is confirmed in future and 
 compared with ${\Delta}M_{B_d}$. Next, let us study 
 ${\Delta}M_D$. In order to estimate the lower limit of 
 ${\Delta}M_D^{SW}$(denoted by $({\Delta}M_D^{SW})_{LL}$) 
 we set ${\mu}_D={\mu}_K$ tentatively in Eq,(24) 
 and obtain
\begin{equation}
 ({\Delta}M_D^{SW})_{LL}=4.67 \times {\Delta}M_K
           = 1.6  \times 10^{-14}\hspace{1cm} {\rm GeV},
 \label{36}
\end{equation}
 where we use $B_D{f}_D^2=(0.19 {\rm GeV})^2$ and Eq.(20), 
 (21), (22), (23a) and (32a). In the same way, assuming 
 ${\mu}_D=1.5 \times {\mu}_K$ for example and 
 using Eq.(24) we have 
\begin{equation}
 {\Delta}M_D^{SW}
  =(2.9 \sim 5.4) \times 10^{-14}\hspace{3.1cm}{\rm GeV},
 \label{37}
\end{equation}
 which is consistent and comparable with Eq.(32b).
 These values are similar to the results 
 by Cheng and Sher[44] and Liu and Wolfenstein[40]. 
\par 
 The study of ${\Delta}M_{B_s}$ is as follows. 
 Both $s$- and $b$-quark in $B_s$-meson are rather massive 
 and then supposing availability of the non-relativistic 
 scheme we have
\begin{equation} 
 {\mu}_{B_s}
    ={m}_s {m}_b/({m}_s+{m}_b)
    =0.19 \hspace{2.7cm}{\rm GeV}, \label{38}
\end{equation}
 where ${m}_s=0.2$ GeV and ${m}_b=4.3$ GeV
 are used.
 If we adopt ${\mu}_K=0.01$ GeV($\simeq{m}_d$) 
 for example we obtain 
 {  
\setcounter{enumi}{\value{equation}}
\addtocounter{enumi}{1}
\setcounter{equation}{0}
\renewcommand{\theequation}{\theenumi\alph{equation}}
\begin{eqnarray}
 {\eta}(B_s) &=& 19.0 {\xi}
\hspace{2cm}{\rm for\hspace{3mm}linear-potential},\\
                    &=& 82.8 {\xi}
\hspace{2cm}{\rm for\hspace{8mm}log-potential},\label{39}
\end{eqnarray}
\setcounter{equation}{\value{enumi}}}
 By using Eq.(20b), (21), (22) and (23a) we have 
\begin{equation} 
     {\Delta}M_{B_s}^{SW}= 
 2(B_{B_s}{f}^2_{B_s}M_{B_s}{\eta}(B_s)/
 B_K{f}^2_KM_K{\eta}(K)){\Delta}M_K^{SW}, \label{40}
\end{equation}
 where factor 2 comes from $n_{B_s}=2$ in Eq.(22).
 Assuming that ${\Delta}M_K^{SW}={\Delta}M_K^{exp}$ 
 (that is, the super-weak exchange saturates the 
 ${\Delta}M_K^{exp}$ bound) and using Eq.(39) we obtain 
\begin{equation}
 {\Delta}M_{B_s}^{SW}=(0.31\sim1.4)\times10^{-11}
 \hspace{3cm}{\rm GeV}, \label{41}
\end{equation}
 where we use $B_{B_s}{f}^2_{B_s}=(0.25{\rm GeV})^2$[42]. 
 From Eq.(32c) and(41) we get 
 {  
\setcounter{enumi}{\value{equation}}
\addtocounter{enumi}{1}
\setcounter{equation}{0}
\renewcommand{\theequation}{\theenumi\alph{equation}}
\begin{eqnarray}
 {\Delta}M_{B_s}^{SW}/{\Delta}M_{B_d}^{SW} &=& 
  10 \sim 50,\\
            x_s={\tau}_{B_s}{\Delta}M_{B_s}&=&
 8 \sim 30, \label{42}
\end{eqnarray}
\setcounter{equation}{\value{enumi}}}
 where we set ${\Delta}M_{B_d}^{SW}={\Delta}M_{B_d}^{exp}$ 
 and use ${\tau}_{B_s}=2.4
         \times10^{12}$ ${\rm GeV}^{-1}$[4].
 Note that the present experimental result is  
 ${\Delta}M_{B_s}^{exp}/{\Delta}M_{B_d}^{exp}>12$[43].
 If we adopt the box diagram calculation in the SM and 
 use Eq.(18) with the unitary assumption of CKM-matrix 
 elements, it is found that[42][43] 
\begin{equation} 
 {\Delta}M_{B_s}^{SD}/{\Delta}M_{B_d}^{SD} 
        =10\sim100. \label{43}
\end{equation}
 Therefore, from the studies of ${\Delta}M_{B_d}$ 
 and ${\Delta}M_{B_s}$ it is difficult to clarify 
 which scheme (SW or SD in the SM) is true, 
 at least until the future experiments confirm 
 the values of $|V_{ts}/V_{td}|$ and ${\Delta}M_{B_s}$. 
\par  
 Finally let us estimate ${\Delta}M_{T_u}^{SW}$ and 
 ${\Delta}M_{T_c}^{SW}$. 
 Setting ${\mu}_{T_u}={\mu}_D$
 (though ${\mu}_{T_u}>{\mu}_D$ in practice)  
 and using Eq.(20b), (21), (22), (23b) and (24) we estimate 
 the lower limit of ${\Delta}M_{T_u}^{SW}$ (denoted by 
 $({\Delta}M_{T_u}^{SW})_{LL}$) as 
\begin{equation} 
      ({\Delta}M_{T_u}^{SW})_{LL}= 
 (B_{T_u}{f}^2_{T_u}M_{T_u}/
  B_{B_d}{f}^2_{B_d}M_{B_d}){\Delta}M_{B_d}^{SW}
      =7.2 \times 10^{-10}\hspace{1cm}{\rm GeV}, \label{44} 
\end{equation}
 where we use $B_{T_u}{f}^2_{T_u}=(1.9{\rm GeV})^2$[3], 
 $M_{B_d}=5.279$ GeV, $M_{T_u}=170$ GeV and set 
 ${\Delta}M_{B_d}^{SW}={\Delta}M_{B_d}^{exp}$ in Eq.(31c). 
 Note that $|{\tilde {\cal M}}_2(T_u)|=
 |{\tilde {\cal M}}_2(B_d)|$ is used in Eq.(44).
 Cheng and Sher's scheme[44] predicts 
 ${\Delta}M_{T_u}\simeq10^{-7}$ GeV which is order $10^3$
 larger than Eq.(44). (In Ref.[44] they estimated
 ${\Delta}M_T\simeq 10^{-10}$ GeV using rather small 
 $t$-quark mass value). 
 For evaluating ${\Delta}M_{T_c}$, we calculate 
\begin{equation}
 {\mu}_{T_c}
    ={m}_c {m}_t/({m}_c+{t}_b)
    =1.34 \hspace{2.7cm}{\rm GeV}, \label{45}
\end{equation}
 where ${m}_c=1.35$ GeV and ${m}_t=170$ GeV
 are used. Then we get from Eq.(24) 
 {  
\setcounter{enumi}{\value{equation}}
\addtocounter{enumi}{1}
\setcounter{equation}{0}
\renewcommand{\theequation}{\theenumi\alph{equation}}
\begin{eqnarray}
 {\eta}(T_c) &=& 134 {\xi}
\hspace{2cm}{\rm for\hspace{3mm}linear-potential},\\
                    &=& 1551 {\xi}
\hspace{1.8cm}{\rm for\hspace{7.5mm}log-potential},
 \label{46}
\end{eqnarray}
\setcounter{equation}{\value{enumi}}}
 where we set ${\mu}_K=0.01$ GeV for example.
 After all with Eq.(46) we obtain
\begin{equation} 
      {\Delta}M_{T_c}^{SW}= 
 2(B_{T_c}{f}^2_{T_c}M_{T_c}/ B_K{f}^2_KM_K)
 ({\eta}(T_c)/{\eta}(K)){\Delta}M_K^{SW}
      =(4\sim47) \times 10^{-8}\hspace{5mm}{\rm GeV},
 \label{47} 
\end{equation}
 where we adopt $n_{T_c}=2$, $B_{T_c}{f}^2_{T_c}=
 (1.9{\rm GeV})^2$[3], $M_{T_u}=171.35$ GeV and 
 ${\Delta}M_K^{SW}={\Delta}M_K^{exp}$. 
 Note that $|{\tilde {\cal M}}_1(T_c)|=
 |{\tilde {\cal M}}_1(K)|$ is used in Eq.(47).
\clearpage
\par
  {\bf b. CP violations in $P^0$-$\overline{P^0}$ mixings}
\newline
\par 
 Presently observed CP violations are only in the Kaon 
 system and are still not inconsistent with the SW-model 
 though there exists the discrepancy between E731 and NA31 
 experiments concerning $Re({\epsilon}^{'}/{\epsilon})_{K}$. 
 If $Re({\epsilon}^{'}/{\epsilon})_{K}=0$ is confirmed 
 by experiments, CKM-mixing matrix has no phase factor[48] 
 and CP violation in $K$-systems can be explained 
 only by the SW-theory. 
 If $Re({\epsilon}^{'}/{\epsilon})_{K}\neq0$ is confirmed 
 by experiments (which implies the existence of 
 CP violation by decay modes), there are several 
 standpoints, e.g., only the standard CKM-theory explains 
 it with (or without) the LD-contributions; 
 the SW-interactions coexist with the SM, etc.. 
 On the other hand, at the present stage no evidence of 
 CP violation in other heavier meson systems is found. 
 As widely discussed, CP asymmetries in 
 $B_s \to {\psi}K_s$ and $B \to 2{\pi}$ may give us the 
 crucial clues[41][51]. 
\par
 In this paper we discuss 
 CP violations by mass-mixings which are assumed 
 to be saturated by the SW-interactions. 
 In the CP conserving limit in the $P^0$-meson systems, 
 $M_{12}(P)$s are supposed to be real positive. 
 Note that $CP|P_H>=-|P_H>$ and 
 $CP|P_L>=|P_L>$ where $H$ $(L)$ means heavy (light). 
 If the CP violating SW-interactions are switched on, 
 $M_{12}(P)$ becomes complex. Following G\'erard and
 Nakada's notation[41][45], we write as 
\begin{equation}
 M_{12}=|M_{12}|\exp(i{\theta}_P), \label{48} 
\end{equation} 
 with 
\begin{equation} 
 \tan{\theta}_P=ImM_{12}(P)/ReM_{12}(P). \label{49}
\end{equation}
 As we assume that the SW-interaction saturates 
 CP violation, we can write 
\begin{equation} 
 Im<\overline{P^0}|H_{SW}^{{\Delta}F=2}|P>=
       ImM_{12}(P). \label{50}
\end{equation} 
 From Eq.(20), (21) and (22) we obtain  
\begin{equation} 
 ImM_{12}(P)={\cal A} \cdot Im{\tilde {\cal M}}_i(P),
 \label{51}
\end{equation}
 where ${\cal A}=
   (1/12\pi^2)B_P{f}^2_PM_P{\eta}(P)$. 
 Therefore the origin of CP violation of $P^0$-meson 
 system is only in ${\tilde {\cal M}_i(P)}$. 
 The Factor ``${\cal A}$'' in Eq.(51) is common also in 
 $ReM_{12}(P)$ and then we have 
\begin{equation} 
  ImM_{12}(P)/ReM_{12}(P)=
     Im{\tilde {\cal M}}_i(P)/Re{\tilde {\cal M}}_i(P).
 \label{52}
\end{equation}          
 If the universality of Eq.(23) is admitted, we obtain 
 {  
\setcounter{enumi}{\value{equation}}
\addtocounter{enumi}{1}
\setcounter{equation}{0}
\renewcommand{\theequation}{\theenumi\alph{equation}}
\begin{eqnarray}
 {\theta}_K &=& {\theta}_D=
 {\theta}_{B_s} = {\theta}_{T_c},\\
 {\theta}_{B_d} &=& {\theta}_{T_u}, \label{53}
\end{eqnarray}
\setcounter{equation}{\value{enumi}}}
 These are the predictions about CP violations from 
 the stand point of our FB-model. Whether 
 ${\theta}_K$ and ${\theta}_{B_d}$ are same or not 
 may depend on whether ${\tilde {\cal M}}_1(K)$ is the same 
 or not as ${\tilde {\cal M}}_2(B_d)$. 
 We know the experimental result as 
\begin{equation} 
 {\theta}_K=(6.5\pm0.2) \times 10^{-3},\label{54}
\end{equation}  
 which is appeared in Ref.[41]. 
 Therefore if this FB-model is admissible, the indirect 
 CP violations of other mesons are also very small and 
 difficult to observe. But as G\'erard and Nakada[41] and 
 Soares and Wolfenstein[51] have pointed out, 
 the measurements of asymmetries of 
 $B \to {\psi}K$ and $2{\pi}$ decays will distinguish the 
 standard CKM-model from the SW-model. 
 For the same purpose we wish to carry out the precise 
 measurements of ${\Delta}M_D$; the dilepton charge 
 asymmetry and also the total charge asymmetry of 
 $D^0$-$\overline{D^0}$ system, which surely discriminate 
 which model is true one. 
 If the future experiments confirm that ${\Delta}M_D \simeq
 10^{-14} \sim 10^{-13}$ GeV 
 and ${\theta}_K \simeq {\theta}_D \simeq {\theta}_{B_s}$, 
 it could be said that the physics in TeV energy region 
 (in which ${\bf y}$-particles play the essential role) 
 totally controls ${\Delta}M_P$ and the indirect 
 CP violations. 
\newline
 {\bf Acknowledgements}
\par 
 We would like to thank to the hospitality at the 
 laboratory of the elementary particle physics 
 in Nagoya University.


\setlength{\unitlength}{1mm}
\begin{picture}(180,270)(15,-20)
\put(0,230){\makebox(30,20){\Huge {Fig.(1)}}\hspace{2cm}
     Subquark line-diagrams of the weak interactions}
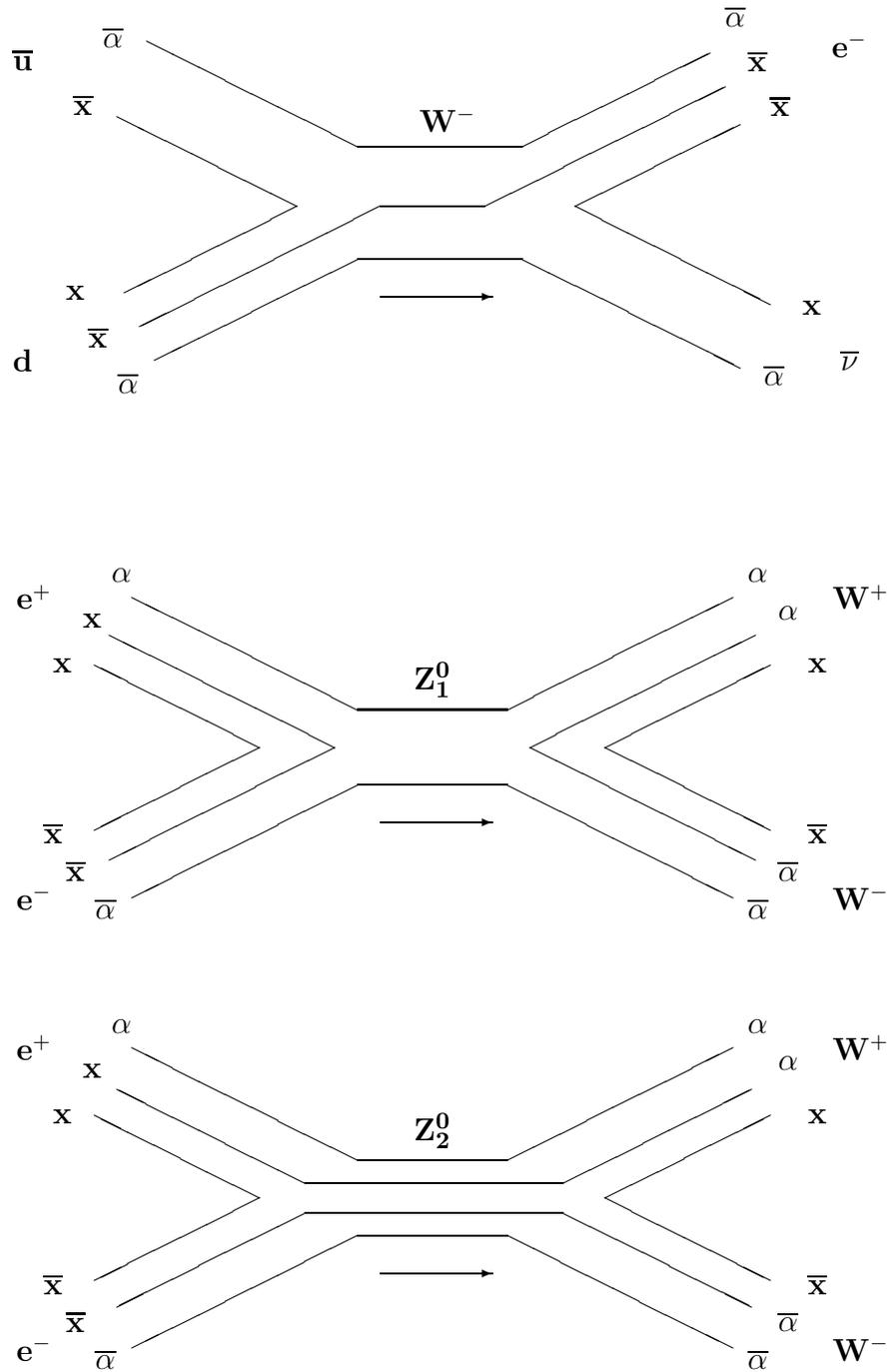
\put(47,216){\makebox(7,7){${\bf \overline{\alpha}}$}}
\put(43,207){\makebox(7,7){${\bf \overline{x}}$}}
\put(35,213){\makebox(7,7){${\bf \overline u}$}}
\put(49,170){\makebox(7,7){${\bf \overline{\alpha}}$}}
\put(45,176){\makebox(7,7){${\bf \overline x}$}}
\put(42,182){\makebox(7,7){${\bf x}$}}
\put(35,173){\makebox(7,7){${\bf d}$}}
\put(132,221){\makebox(7,7)[bl]{${\bf \overline{\alpha}}$}}
\put(135,215){\makebox(7,7)[bl]{${\bf \overline x}$}}
\put(138,209){\makebox(7,7)[bl]{${\bf \overline x}$}}
\put(145,215){\makebox(7,7){${\bf e^-}$}}
\put(140,180){\makebox(7,7){${\bf x}$}}
\put(135,171){\makebox(7,7){${\bf \overline{\alpha}}$}}
\put(145,173){\makebox(7,7){${\bf \overline{\nu}}$}}
\put(90,207){\makebox(10,10)[b]{${\bf W^-}$}}
\put(83,205){\line(-2,1){28}}
\put(75,197){\line(-2,1){24}}
\put(75,197){\line(-2,-1){23}}
\put(86,197){\line(-2,-1){32}}
\put(83,190){\line(-2,-1){27}}
\put(86,185){\vector(1,0){15}}
\put(83,205){\line(1,0){22}}
\put(86,197){\line(1,0){14}}
\put(83,190){\line(1,0){22}}
\put(105,205){\line(2,1){25}}
\put(100,197){\line(2,1){32}}
\put(112,197){\line(2,1){22}}
\put(112,197){\line(2,-1){26}}
\put(105,190){\line(2,-1){29}}
\put(46,147){\makebox(7,7)[br]{${\bf {\alpha}}$}}
\put(42,141){\makebox(7,7)[br]{${\bf x}$}}
\put(38,135){\makebox(7,7)[br]{${\bf x}$}}
\put(39,110){\makebox(7,7){${\bf \overline{x}}$}}
\put(42,105){\makebox(7,7){${\bf \overline x}$}}
\put(46,100){\makebox(7,7){${\bf \overline{\alpha}}$}}
\put(135,147){\makebox(7,7)[bl]{${\bf \alpha}$}}
\put(139,142){\makebox(7,7)[bl]{${\bf \alpha}$}}
\put(143,135){\makebox(7,7)[bl]{${\bf x}$}}
\put(143,110){\makebox(7,7)[l]{${\bf \overline{x}}$}}
\put(139,105){\makebox(7,7)[l]{${\bf \overline{\alpha}}$}}
\put(135,100){\makebox(7,7)[l]{${\bf \overline{\alpha}}$}}
\put(35,140){\makebox(10,10){${\bf e^+}$}}
\put(35,100){\makebox(10,10){${\bf e^-}$}}
\put(145,140){\makebox(10,10){${\bf W^+}$}}
\put(145,100){\makebox(10,10){${\bf W^-}$}}
\put(88,129){\makebox(10,10){${\bf Z_1^0}$}}
\put(83,130){\line(-2,1){30}}
\put(80,125){\line(-2,1){30}}
\put(70,125){\line(-2,1){22}}
\put(70,125){\line(-2,-1){22}}
\put(80,125){\line(-2,-1){30}}
\put(86,115){\vector(1,0){15}}
\put(83,120){\line(-2,-1){30}}
\put(83,130){\line(1,0){20}}
\put(83,120){\line(1,0){20}}
\put(103,130){\line(2,1){30}}
\put(106,125){\line(2,1){30}}
\put(116,125){\line(2,1){22}}
\put(116,125){\line(2,-1){22}}
\put(106,125){\line(2,-1){30}}
\put(103,120){\line(2,-1){30}}
\put(46,87){\makebox(7,7)[br]{${\bf {\alpha}}$}}
\put(42,81){\makebox(7,7)[br]{${\bf x}$}}
\put(38,75){\makebox(7,7)[br]{${\bf x}$}}
\put(39,50){\makebox(7,7){${\bf \overline{x}}$}}
\put(42,45){\makebox(7,7){${\bf \overline x}$}}
\put(46,40){\makebox(7,7){${\bf \overline{\alpha}}$}}
\put(135,87){\makebox(7,7)[bl]{${\bf \alpha}$}}
\put(139,82){\makebox(7,7)[bl]{${\bf \alpha}$}}
\put(143,75){\makebox(7,7)[bl]{${\bf x}$}}
\put(143,50){\makebox(7,7)[l]{${\bf \overline{x}}$}}
\put(139,45){\makebox(7,7)[l]{${\bf \overline{\alpha}}$}}
\put(135,40){\makebox(7,7)[l]{${\bf \overline{\alpha}}$}}
\put(35,80){\makebox(10,10){${\bf e^+}$}}
\put(35,40){\makebox(10,10){${\bf e^-}$}}
\put(145,80){\makebox(10,10){${\bf W^+}$}}
\put(145,40){\makebox(10,10){${\bf W^-}$}}
\put(88,69){\makebox(10,10){${\bf Z_2^0}$}}
\put(83,70){\line(-2,1){30}}
\put(76,67){\line(-2,1){25}}
\put(70,65){\line(-2,1){22}}
\put(70,65){\line(-2,-1){22}}
\put(76,63){\line(-2,-1){25}}
\put(86,55){\vector(1,0){15}}
\put(83,60){\line(-2,-1){30}}
\put(83,70){\line(1,0){20}}
\put(76,67){\line(1,0){34.5}}
\put(76,63){\line(1,0){34.5}}
\put(83,60){\line(1,0){20}}
\put(103,70){\line(2,1){30}}
\put(110.5,67){\line(2,1){25}}
\put(116,65){\line(2,1){22}}
\put(116,65){\line(2,-1){22}}
\put(110.5,63){\line(2,-1){25}}
\put(103,60){\line(2,-1){30}}
\end{picture}
\setlength{\unitlength}{1mm}
\begin{picture}(180,250)(15,-20)
\put(0,200){\makebox(30,20){\Huge {Fig.(2)}}\hspace{0.3cm}
  Schematic illustrations of $P^0$-$\overline{P^0}$ mixings
  by ${\bf y}$-exchange interactions}
\put(140,155){\makebox(30,30)[l]{${\overline{K^0}}
             \hspace{1mm}({\overline{D^0}})$}}
\put(0,155){\makebox(30,30)[r]{${K^0}
             \hspace{1mm}({D^0})$}}
\put(30,182){\makebox(15,10)[r]{${\bf {\overline{s}}
             \hspace{1mm}(c)}$}}
\put(60,190){\line(1,0){60}}
\put(67,191){\makebox(50,10)[b]{${\bf {\alpha x \overline{x}
                                 \hspace{1mm}(\alpha x)}}$}}
\put(130,182){\makebox(15,10)[l]{${\bf {\overline {d}}
             \hspace{1mm}(u)}$}}
\put(50,180){\makebox(7,7)[r]{${\bf y}$}}
\put(60,183){\line(1,0){30}}
\put(90,160){\line(0,1){23}}
\put(90,160){\line(1,0){30}}
\put(123,157){\makebox(7,7)[l]{${\bf y}$}}
\put(30,150){\makebox(15,10)[r]{${\bf {d\hspace{1mm}
             (\overline{u})}}$}}
\put(60,153){\line(1,0){60}}
\put(130,151){\makebox(15,10)[l]{${\bf {s\hspace{1mm}
              (\overline{c})}}$}}
\put(67,154){\makebox(50,10)[b]{${\bf {\overline{\alpha} 
    \overline{x} x \hspace{1mm}(\overline{\alpha} 
    \overline{x})}}$}}
\put(140,95){\makebox(30,30)[l]{${\overline{B_s^0}}
             \hspace{1mm}({\overline{T_c^0}})$}}
\put(0,95){\makebox(30,30)[r]{${B_s^0}
             \hspace{1mm}({T_c^0})$}}
\put(30,120){\makebox(15,10)[r]{${\bf {\overline{b}}
             \hspace{1mm}(t)}$}}
\put(60,130){\line(1,0){60}}
\put(67,131){\makebox(50,10)[b]{
            ${\bf {\alpha x \overline{x}
                                \hspace{1mm}(\alpha x)}}$}}
\put(130,123){\makebox(15,10)[l]{${\bf {\overline {s}}
             \hspace{1mm}(c)}$}}
\put(50,122){\makebox(7,7)[r]{${\bf y}$}}
\put(60,125){\line(1,0){60}}
\put(123,122){\makebox(7,7)[l]{${\bf y}$}}
\put(50,116){\makebox(7,7)[r]{${\bf y}$}}
\put(60,120){\line(1,0){30}}
\put(90,100){\line(0,1){20}}
\put(90,100){\line(1,0){30}}
\put(123,97){\makebox(7,7)[l]{${\bf y}$}}
\put(50,91){\makebox(7,7)[r]{${\bf y}$}}
\put(60,95){\line(1,0){60}}
\put(123,91){\makebox(7,7)[l]{${\bf y}$}}
\put(30,87){\makebox(15,10)[r]{${\bf {s
           \hspace{1mm}(\overline{c})}}$}}
\put(60,90){\line(1,0){60}}
\put(67,91){\makebox(50,10)[b]{${\bf {\overline{\alpha} 
    \overline{x} x \hspace{1mm}(\overline{\alpha} 
    \overline{x})}}$}}
\put(130,90){\makebox(15,10)[l]{${\bf {b\hspace{1mm}
             (\overline{t})}}$}}
\put(140,35){\makebox(30,30)[l]{${\overline{B_d^0}}
             \hspace{1mm}({\overline{T_u^0}})$}}
\put(0,35){\makebox(30,30)[r]{${B_d^0}
             \hspace{1mm}({T_u^0})$}}
\put(30,60){\makebox(15,10)[r]{${\bf {\overline{b}}
             \hspace{1mm}(t)}$}}
\put(60,70){\line(1,0){60}}
\put(67,71){\makebox(50,10)[b]{${\bf {\alpha x \overline{x}
                                \hspace{1mm}(\alpha x)}}$}}
\put(130,63){\makebox(15,10)[l]{${\bf {\overline {d}}
             \hspace{1mm}(u)}$}}
\put(50,62){\makebox(7,7)[r]{${\bf y}$}}
\put(60,65){\line(1,0){32}}
\put(92,40){\line(0,1){25}}
\put(50,56){\makebox(7,7)[r]{${\bf y}$}}
\put(60,60){\line(1,0){28}}
\put(88,35){\line(0,1){25}}
\put(92,40){\line(1,0){28}}
\put(123,37){\makebox(7,7)[l]{${\bf y}$}}
\put(88,35){\line(1,0){32}}
\put(123,31){\makebox(7,7)[l]{${\bf y}$}}
\put(30,27){\makebox(15,10)[r]{${\bf {d
           \hspace{1mm}(\overline{u})}}$}}
\put(60,30){\line(1,0){60}}
\put(67,31){\makebox(50,10)[b]{${\bf {\overline{\alpha} 
           \overline{x} x \hspace{1mm}
           (\overline{\alpha} \overline{x})}}$}}
\put(130,30){\makebox(15,10)[l]{${\bf {b\hspace{1mm}
             (\overline{t})}}$}}
\end{picture}

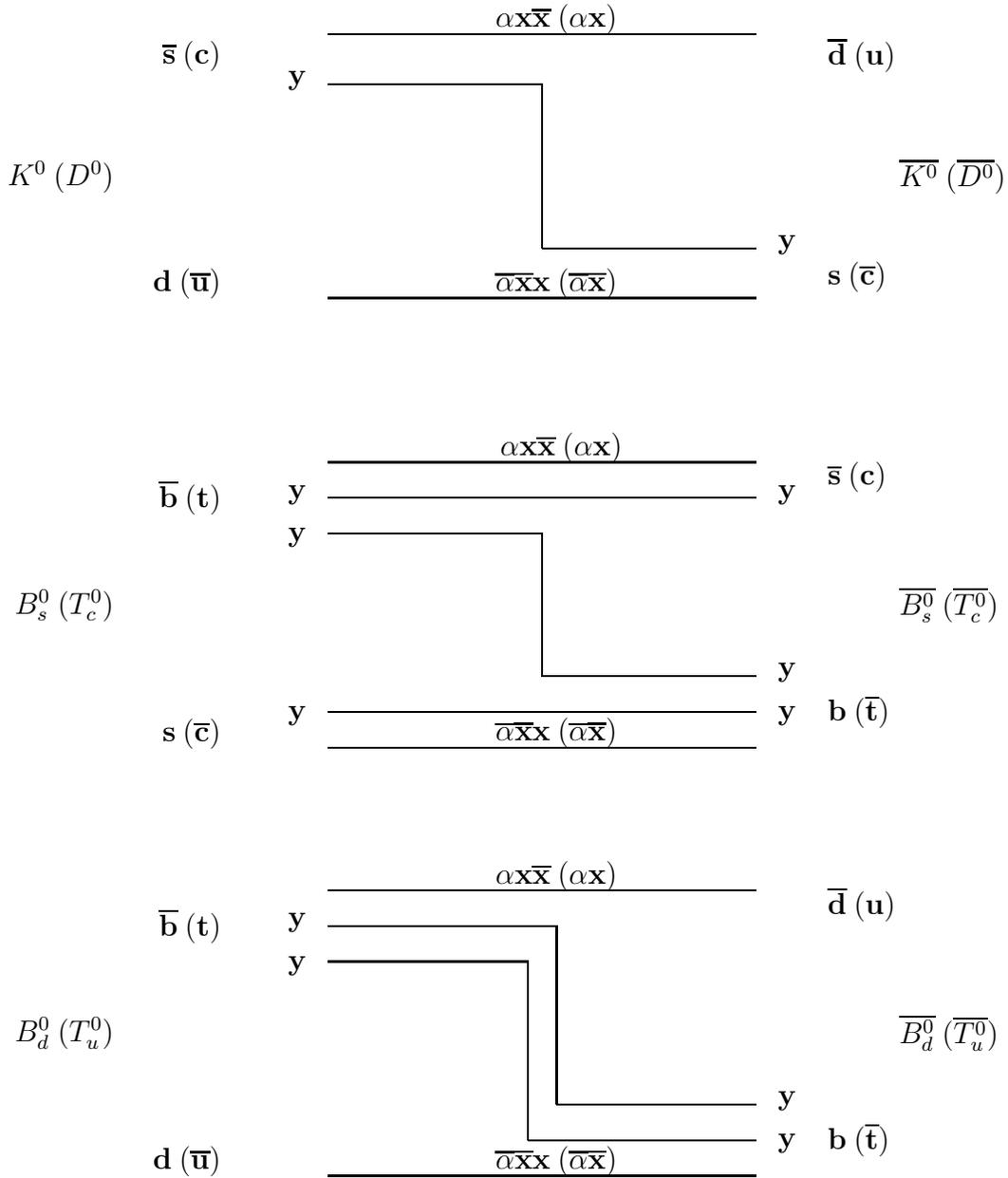
\end{document}